\documentclass{aastex}
\usepackage{emulateapj5}
\input{psfig.tex}

\def\chandra{{\it Chandra}} 
 
\def\asca{{\it ASCA}} 
\def\hst{{\it HST}} 
\def\sirtf{{\it SIRTF}}

\def\rosat{{\it ROSAT}} 
\def\ein{{\it Einstein}}

\def\vla{{\it VLA}}

\def\flux{erg cm$^{-2}$ s$^{-1}$}

\def\arcsec{$^{\prime\prime}$}
\def\deg{$^{\circ}$}

\def\ltsima{$\; \buildrel < \over \sim \;$}
\def\simlt{\lower.5ex\hbox{\ltsima}} 
\def\gtsima{$\; \buildrel > \over \sim \;$}
\def\simgt{\lower.5ex\hbox{\gtsima}} 

\begin{document}

\title{A survey of extended radio jets with Chandra and HST}

\normalsize
\author{Rita M. Sambruna} \affil{George Mason University, Dept. of
Physics and Astronomy and School of Computational Sciences, MS 3F3,
4400 University Drive, Fairfax, VA 22030 (rms@physics.gmu.edu)}

\author{Jessica K. Gambill} \affil{George Mason University, School of
Computational Sciences, MS 5C3, 4400 University Drive, Fairfax, VA 22030}

\author{L. Maraschi, F. Tavecchio, and R. Cerutti} \affil{Osservatorio Astronomico
di Brera, via Brera 28, 20121 Milano, Italy}

\author{C. C. Cheung} \affil{Brandeis University, Department of
Physics, MS 057, Waltham, MA 02454}

\author{C. Megan Urry} \affil{Yale University, Dept. of Astronomy, New
Haven, CT 06520}

\author{G. Chartas} \affil{The Pennsylvania State University, Dept. of
Astronomy and Astrophysics, 525 Davey Lab, State College, PA 16802}

\begin{abstract}
We present the results from an X-ray and optical survey of a sample of
17 radio jets in Active Galactic Nuclei performed with \chandra\ and
\hst. The sample was selected from the radio and is unbiased toward
detection at shorter wavelengths, but preferentially it includes beamed
sources. We find that X-ray emission is common on kpc-scales, with
over half (10/17) radio jets exhibiting at least one X-ray knot on the
\chandra\ images. A similar detection rate is found for the optical emission,
although not all X-ray knots have optical counterparts, and
vice-versa. The distributions of the radio-to-X-ray and
radio-to-optical spectral indices, $\alpha_{rx}$ and $\alpha_{ro}$,
for the detected jets are similar to the limits for the
non-detections, suggesting all bright radio jets have X-ray
counterparts which will be visible in longer observations. Comparing
the radio and X-ray morphologies shows that the majority of the X-ray
jets have structures that closely map the radio. Analysis of the
Spectral Energy Distributions of the jet knots suggest the knots in
which the X-ray and radio morphologies track each other produce X-rays
by inverse Compton (IC) scattering of the Cosmic Microwave Background
(IC/CMB). The remaining knots produce X-rays by the synchrotron
process. Spectral changes are detected along the jets, with the ratio
of the X-ray-to-radio and optical-to-radio flux densities decreasing
from the inner to the outer regions. This suggests the presence of an
additional contribution to the X-ray flux in the jet's inner part,
either from synchrotron or IC of the stellar light. Alternatively, in
a pure IC/CMB scenario, the plasma decelerates as it flows from the
inner to the outer regions. Finally, the X-ray spectral indices for
the brightest knots are flat (photon index $\Gamma_X \sim 1.5$),
indicating that the bulk of the luminosity of the jets is emitted at
GeV energies, and raising the interesting possibility of future
detections with GLAST.

{\sl Subject Headings:}{Galaxies: active --- galaxies: jets ---
(galaxies:) quasars: individual --- X-rays: galaxies}

\end{abstract} 

\section{Introduction}

Before the advent of the \chandra\ X-ray Observatory in 1999, X-ray
emission from kiloparsec-scale jets in Active Galactic Nuclei (AGNs)
was virtually unknown. Only a few X-ray jets were known from the
previous generation of X-ray telescopes (mainly \ein\ and \rosat),
including the optical jets of M87 and 3C~273 (see Sparks, Biretta, \&
Macchetto 1997 and references therein) and the radio jets of
Centaurus~A and NGC~6251 (Feigelson et al. 1981, Mack et al. 1997). A
few more jets were known to emit in the optical, an extension of the
radio synchrotron emission (e.g., Scarpa \& Urry 2002; Sparks et
al. 1994). With only about a dozen of the hundreds of the known radio
jets detected at both optical and X-rays, high-energy emission was
viewed as an exotic phenomenon.

The launch of \chandra, with its improved resolution and sensitivity
over previous X-ray satellites, opened a new window for the study of
jets, allowing for the first time imaging spectroscopy studies of
these structures. The first \chandra\ light, the distant quasar PKS
0637$-$752, surprisingly showed a bright, kpc-scale X-ray jet (Chartas
et al. 2000), with only a weak optical counterpart in archival \hst\
data (Schwartz et al. 2000). Indeed, intense X-ray emission had been 
detected from radio jets that were not known to have optical
counterparts (e.g., Pictor A; Wilson, Young, \& Shopbell 2001), and
studies of previously known synchrotron optical jets showed that the
X-rays do not always lie on the extrapolation of the optical emission,
challenging simple synchrotron models (e.g., M87, Wilson \& Yang 2002;
3C~273, Sambruna et al. 2001). It quickly became clear that the X-ray
emission from jets is very complex. 

The bright X-ray flux in PKS 0637--752 could not be well explained as
synchrotron emission or synchrotron-self Compton (Schwartz et
al. 2000), but was instead attributed to inverse Compton (IC)
scattering of Cosmic Microwave Background (CMB) photons by
relativistic electrons in the jet, with jet Lorentz factors $\Gamma
\sim 10$ (Tavecchio et al. 2000; Celotti et al. 2001).  The IC/CMB
model can also account for the X-ray emission from some knots in the
3C~273 jet (Sambruna et al. 2001) and for the distant jet in the
gravitationally lensed quasar Q0957+561 (Chartas et al. 2002). Because
of the $(1+z)^4$ dependence of the CMB density, compensating the
surface brightness dimming, high-redshift jets should in fact be as
bright at X-rays as low-z jets (Schwartz 2002).

Thus, early \chandra\ results showed that X-ray emission from
kpc-scale jets may be more common than previously thought on the basis
of lower sensitivity detectors. Moreover, a reanalysis of the \hst\
archival data prompted by the \chandra\ results yielded a few more
optical detections, showing that optical emission from jet may also be
common (e.g., Cheung 2002, Schwartz et al. 2000). Additionally,
\chandra\ is now showing that the terminal lobes and hotspots in
powerful FRII jets emit at X-rays, although the origin of the X-ray
flux from these structures is still a matter of debate (e.g.,
Hardcastle et al. 2002a).

We initiated a systematic study of the optical and X-ray emission from
extended jets in AGN, in order to address their physical properties
(magnetic fields, particle energy distributions, plasma speeds and
power), surveying a well-defined sample of radio jets with
\chandra\ and \hst\ with relatively short exposures. 
The results for the first six targets observed were presented in
Sambruna et al. (2002; hereafter Paper I). Here we report the X-ray
results for the whole sample of 17 jets; the six targets of Paper I
were reanalyzed in light of the improved ACIS calibration and the
\hst\ results. We determine jet detection rates, compare the multiwavelength
morphologies, and discuss emission mechanisms and outstanding
questions regarding extended jet emission. We reanalyzed archival
\vla\ radio data. The radio and optical data on the jets are presented
in detail in Cheung et al. (2004) and Urry et al. (2004),
respectively, while the unresolved X-ray emission from the nuclei of
the sample objects is discussed in Gambill et al. (2003).  A
multiwavelength study of the hotspots and lobe features will be
presented in a future paper.

The plan of this paper is as follows. In \S~2 we review the sample
selection criteria, and in \S~3 the observations and the data
analysis. In \S~4 we present the results, and in \S~5 we discuss the
origin of the X-ray emission from the kpc-scale jets and their
physical properties. Summary and conclusions are reported in
\S~6. Throughout this work, H$_0=75$ km s$^{-1}$ Mpc$^{-1}$ and
$q_0=0.5$ are adopted.

\section{The Jet Sample}

The targets of the program were selected according to radio-driven
criteria, without knowledge of the optical and X-ray emission
properties. In this sense, the survey is unbiased toward emission at
the longer wavelengths. However, the survey is biased toward beaming
because of the radio selection criteria (see below).

The survey sample was chosen from the list of known radio jets of
Bridle \& Perley (1984) and Liu \& Xie (1992), according to the following 
criteria. (1) 
The radio jet has surface brightness $S_{1.4~GHz}$ \simgt
5mJy/arcsec$^2$ at $>$ 3\arcsec\ from the nucleus, i.e., long enough
and bright enough to be detected in reasonable \chandra\ and
\hst\ exposures for average values of the radio-to-X-ray and
radio-to-optical spectral indices, $\alpha_{rx} \sim 0.8$ and
$\alpha_{ro} \sim 0.8$. (2) High-resolution (1\arcsec\ or better)
published or archival radio maps show that at least one bright (\simgt
5 mJy) radio knot is present at $>$ 3\arcsec\ from the nucleus. The
resulting sample of 17 radio jets spans a range of redshifts, core and
extended radio powers, and classifications: 16 sources are classified
as quasars and are hosted by ellipticals classified as powerful
Fanaroff-Riley II (FRII; Fanaroff \& Riley 1974) radio galaxies, while
the low-power, nearby source 0836+299 is hosted by a radio galaxy with
FRI morphology (most apparent on low frequency maps; van Breugel et
al. 1986). 

The targets are listed in Table 1 together with their basic
properties: redshift (column 3), scale conversion (projected size;
column 4), Galactic column density (column 5), radio classification of
the core (column 6), core radio power at 5 GHz (columns 7), ratio of
core to extended radio power, $R_i$ (column 8), and limits on the
ratio of the jet-to-counterjet fluxes, $J$ (column 9). Measurements of
$R_i$ are traditionally used to separate lobe-dominated ($R_i < 1$)
from core-dominated ($R_i > 1$) radio sources, although the measured
values are sensitive to the observing frequency (e.g., Orr \& Browne
1982). At lower observing frequencies, where the steep spectrum lobes
will become more prominent, $R_i$ values may decrease.  Indeed, 1.65
GHz observations (published, and our own analysis of archival data;
Cheung et al. 2004) indicate that 3 of the 11 source classified as
Flat Spectrum Radio Quasars (FSRQs) based on the 5 GHz data, have
$R_i$ values less than unity at 1.65 GHz, indicating that they are
more Steep Spectrum Radio Quasar (SSRQ)-like or simply less beamed
FSRQs.

In general, all the sources of the sample show one-sided jets,
indicating substantial beaming. In addition, most of the sources
exhibit superluminal motion at VLBI scales with large apparent speeds,
again indicating beaming. Indeed, most sources are classified as
FSRQs. The core-to-lobe ratio, $R_i$, can be used as a beaming
indicator with scatter depending on the scatter in intrinsic
core-to-extended luminosity.

During \chandra\ Cycle 2, we were awarded 10 kilo second (ks)
exposures for 16 of the 17 sources with \chandra\ ACIS-S, and one
orbit per target with
\hst\ for all 17 targets.  The X-ray flux limit for a 2$\sigma$
detection in a typical 10-ks ACIS-S observation is $F_{0.5-8 keV} \sim
2 \times 10^{-15}$ \flux.  The \chandra\ observations of the 17th
source, 3C 207, was awarded to other investigators (Brunetti et
al. 2002), with a 37 ks exposure.  Here we include the archival
\chandra\ data of 3C~207, which we reanalyzed to ensure uniformity
with the remaining sources and in light of the improved
calibration. The six sources previously discussed in Paper I
(0723+679, 1055+018, 1136$-$135, 1150+497, 1354+195, and 2251+134)
have been also reanalyzed and are included here. It is important to
note that the short \chandra\ and \hst\ exposures of this survey were
designed to detect the X-ray/optical counterpart of the radio jets,
leaving detailed studies of jet morphologies and spectra for deeper
follow-up observations.

\section{Data Acquisition, Reduction, and Analysis}

\subsection{Chandra Data} 

The \chandra\ observations were performed with ACIS-S with the source
at the aimpoint of the S3 chip.  Since we expected bright X-ray cores,
we used $\frac{1}{8}$ subarray mode with an effective frame time of
0.4 s, to reduce the effect of pileup of the nucleus.  In addition,
for each source, a range of roll angles was specified in order to
locate the jet away from the charge transfer trail of the nucleus and
avoid flux contamination. Nevertheless, the cores of 10/17 sources
were affected by pileup (Gambill et al. 2003).  The fraction of core
pileup is \gtsima5\%, according to the measured counts per frame
(\gtsima 0.26 counts per frame). The distance from the core at which
the PSF wings can contaminate the jet varies with the level of core
pileup in the specific targets. The sources with the largest amount of
core pileup are 0405$-$123 and 1150+497, with 0.5 c/frame.

The \chandra\ data were reduced following standard screening criteria
and using the latest calibration files provided by the \chandra\ X-ray
Center. The latest version of the reduction software \verb+CIAO+
v. 2.3  was used. Pixel randomization was removed, and only events
for \asca\ grades 0, 2--4, and 6 and in the energy range 0.5--8 keV, in
which the background is negligible and the ACIS-S calibration best
known, were retained.  We also checked that no flaring background
events occurred during the observations.  After screening, the
effective exposure times range between 7 and 10 ks, with one object
(0838+133) observed for much longer, 23.9 ks (Table~2). Note that
these exposures are slightly different from those in Paper I and
Gambill et al. (2003), due to the revised ACIS-S calibration. 

Several counterparts to radio jet features were detected in the X-ray
band in our \chandra\ images (Figures 1, 2, and Table 2).  X-ray
counts from each knot were extracted in a circular region of radius
1\arcsec, with the background estimated in a circular region of radius
10\arcsec\ at a nearby position free of spurious X-ray sources.  For a
point source, the extraction radius of 1\arcsec\ encircles
\gtsima90\% of the flux density at 1 keV, based on the ACIS-S encircled energy
fraction (Fig.6.3 in the \chandra\ Proposer Observatory Guide). We
corrected the X-ray fluxes for this effect. We determined corrections
to the ancillary response files by simulating the spectra of point
sources at the locations of the images within the apertures used in
our analysis to correct for the remaining flux.  For our simulations
we used
\verb+XSPEC+ to generate the source spectra and the ray-tracing tool
\verb+MARX+ to model the dependence of photon scattering with energy.

However, in two instances, a circular extraction region of radius
1.5\arcsec\ was used instead of the standard 1.0\arcsec\ region.  A
larger region was used when the lobe or counterlobe emission evidently
spread beyond the 1.0\arcsec\ aperture, since lobe emission is more
diffuse than the concentrated emission of the knots in lower frequency
observations. When two optical knots were observed with \hst, but were
unresolved in the \chandra\ image, a larger region was applied for the
extraction of X-rays since it could contain all of the emission from
the two optical knots.

In several cases, optical knots were detected with \hst\ at \ltsima
1--2\arcsec\ from the core, where a significant contribution from the
PSF wings could be present. Inspection of the X-ray maps showed an
excess of the flux above the PSF wings at the position of the inner
knots.  To derive the net X-ray flux for all the inner knots, we
adopted the following procedures: (1) For knots detected with \hst\ at
\ltsima1.5\arcsec\ from the core, we extracted the counts from a
circular aperture of radius 0.5\arcsec\ centered on the optical
position of the knot. (2) For knots detected with \hst\ between
1.5--2.2\arcsec\ from the core, we extracted the counts from a
circular aperture of radius 1.0\arcsec\ centered on the optical
position of the knot.  To subtract the contribution of the PSF, the
background was evaluated in a circular region with the same radius and
distance from the core used for extraction (0.5\arcsec\ or 1.0\arcsec\
depending on the distance from the core), but at three different
azimuth angles, chosen so as to avoid possible contamination from
pileup and counter-features.

We find that the net (background-subtracted) counts vary by as much as
30\% depending on the background region.  This large uncertainty
derived from background position in the region of the nucleus suggests
local non-uniformities of the core PSF, attributable to the significant
effect of pileup on the core PSF for some sources (Gambill et
al. 2003). Given the large excursion of values, the counts reported in
Tables 2 and 3 for these inner knots (generally coincident with knot
A) are simple average values, while the uncertainties are their
standard deviations. Thus, the values for net X-ray counts represent
rather conservative estimates for the X-ray fluxes of these knots.

The net X-ray count rates or upper limits to them are listed in Tables
2 and 3.  Following Paper I, the count rates and relative
uncertainties are calculated according to the following steps. Let $S$
be the total (before background subtraction) counts in the source area
$A_S$, and $b$ the background counts in the area $A_B$. The background
counts rescaled to the source region are $B=b A_s/A_B$.  The net
counts of the source are thus $N = S - B$.  The uncertainties of the
net X-ray counts, $\sigma_N$, were calculated according to the formula
$\sigma_N=[(\sigma_S)^2 + (\sigma_B)^2]^{1/2}$, where $\sigma_S$ and
$\sigma_B$ are the uncertainties of the source and background,
respectively.  The latter were calculated following Gehrels (1986), as
appropriate in a regime of low counts: $\sigma_S=1+(S+0.75)^{1/2}$ and
$\sigma_B = [1+(b+0.75)^{1/2}](A_S/A_B)$.

We define as detections only the knots/hotspots for which the
significance of the X-ray counts is at least $2\sigma$. When the
detected X-ray counts are significant at less than $2\sigma$, an upper
limit is given in Table 3, as a 1$\sigma$ upper limit (column 6) and
as a count rate (column 7). The upper limits were extracted in a
1\arcsec\ radius region centered on the radio position of the knot,
with the background estimated as discussed above for the case of the
detections.

Upper limits are also given in Table 3 for all the radio features
which were detected in the optical in our \hst\ images, but which do
not have an X-ray counterpart. Moreover, in some cases faint X-ray
counts in excess of the background were detected nearby, but not quite
coincident with, a radio feature. As it is unclear whether these
excess counts represent a true detection or rather a background
fluctuation or even a foreground source, conservatively we give an
upper limit to the X-ray count rate in Table 3. These cases include
0802+103, 0836+299 knot A, 1055+018, 1741+279 knot A and hotspot B;
each is labeled with a question mark in Table 3.

For the knots in Table 2a with 40 counts or more we extracted X-ray
spectra in an aperture of radius 1\arcsec\ centered on the knot (the
same region used to extract the counts). The spectra were rebinned in
order to have at least 5 counts in each new bin, and fitted with
\verb+XSPEC+ v.11.2 in the energy range 0.5--8 keV, where the
calibration is best known and the background negligible. The
C-statistics, appropriate for low signal-to-noise ratio data, were
used to derive the spectral parameters. The background-subtracted
spectra were fitted with a single power law model with fixed Galactic
column density, using the Morrison \& McCammon (1983) cross section
and solar abundances.  We report the spectral index in column 10 of
Table~2a. Errors are 90\% confidence for one parameter of interest.

\subsection{HST Data} 

The results from our \hst\ imaging are described in detail in Urry et
al.  (2004).  In short, we imaged all 17 objects with the STIS CCD in
\verb+CLEAR+ (unfiltered) imaging mode, which is more sensitive than
the \hst\ WFPC2 camera for detecting faint point sources. The purpose
of these exposures was simply to identify optical emission associated
with the jets, allowing for more color-sensitive, deeper, follow-up
studies later. The observations were carried out from November 2000 to
February 2001, with 2251+134 observed in August 2001, and in general
were not simultaneous to the \chandra\ exposures. The one orbit per
target awarded amounted to an average of $\sim$ 2,700 sec of total
integration time from a series of shorter exposures for 16 of the
objects. The quasar 0723+679 was a continuous viewing zone target, so
we were able to obtain a longer sequence of exposures totaling 4,680
seconds. The individual exposures were stacked and cosmic rays removed
using the \verb+CRREJ+ task in IRAF.  Optical emission from the radio
jets was identified via digital overlays with our high resolution
\vla\ images. Count rates for the detected optical jet features were
converted to flux densities utilizing the inverse sensitivity
measurements contained in the \verb+PHOTFLAM+ keyword inside the image
headers which approximately gives $1.08 \times 10^{-7}$ (Jy $\times$
count rate$^{-1}$) at a pivot wavelength of 5852
\AA. The fairly uniform quality of the 17 datasets allows us to
estimate conservatively that the point source sensitivity limit in
each image is approximately 0.06 $\mu$Jy (3$\sigma$), judging by the
fact that the weakest bona fide optical detections were of order 0.02
$\mu$Jy (in 1040+123, 1150+497, and 1642+690; see Urry et al. 2004).
It is important to stress that this limit is applicable only to point
sources and presumes that the non-detected optical counterparts would
appear as unresolved features in deeper \hst\ images. Specifically,
this limit does not apply to diffuse features.

Optical flux densities for the Spectral Energy Distributions described
in \S~5 were extracted as follows. The optical fluxes were obtained by
summing up the contributions of the \hst\ detected features which laid
within the specified X-ray apertures (usually, only one optical
knot). For the features at small angular separations from the nucleus
which were detected in our \hst\ data (\ltsima 1.5\arcsec; marked with
a footnote in Tables 2 and 3), the corresponding radio and optical
flux densities are quoted directly from Urry et al. (2004). The radio
and optical images achieved comparable resolution so we were able to
uniquely match the optically detected features with peaks in the radio
jet. The Urry et al. (2004) radio fluxes were therefore measured by
fitting the radio jet knots with elliptical Gaussian components in the
(u,v) plane using the \verb+DIFMAP+ modelfit program. Since the radio
knots tended to be smaller than 1\arcsec\ in size, fluxes measured in
this manner will give generally smaller fluxes than those obtained
with the larger apertures we used to measure radio fluxes for the more
distant X-ray emitting features discussed above. Care should be taken
in comparing the radio fluxes in this paper with those presented in
Urry et al. (2004). Because the optical knots were often faint
point-like sources which appear over a relatively large background in
the \hst\ images, optical fluxes were obtained by measuring count
rates in successively increasing circular apertures centered on the
optical peak until a plateau was reached -- this mimics an infinite
radius aperture.

\subsection{Archival Radio Data} 

Our complete analysis of the radio images along with polarization
information, will be presented in detail in a forthcoming paper
(Cheung et al. 2004). In summary, {\it Very Large Array} (Thompson et
al. 1980) data at 5 GHz was gathered from the NRAO\footnote{The
National Radio Astronomy Observatory is a facility of the National
Science Foundation operated under cooperative agreement by Associated
Universities, Inc.} archive for all 17 targets. Most of the
observations utilized the \vla\ in its highest resolution
A-configuration which gives better than 0.5\arcsec\ resolution at this
frequency. For the larger sources, we used data from the
B-configuration (0405$-$123, 1928+738), or combined data from both the
A- and B-configurations (1055+018, 1354+195) so the resultant images
achieved resolutions of order 1\arcsec. The quality of the images is
not uniform due to the fact that the data were obtained from separate
programs with widely different integration times so care must be taken
in inspecting the radio-derived parameters in Table~1. The most
shallow image is the 6-min snapshot of 2251+134; the best images are
dedicated full-track observations of 0836+299 and 1040+123 where
almost 3 hrs of data were obtained.  The final images presented in
Figures 1 and 2 are restored with circular beams and are shown at full
resolution in order to display details in the radio jets not obvious
in the \chandra\ images, which have 0.86\arcsec\ resolution.

The basic calibration was performed in \verb+AIPS+ (Bridle \& Greisen
1994b) and then outputted to the Caltech \verb+DIFMAP+ package
(Shepherd, Pearson, \& Taylor 1994) for self-calibration and
imaging. Much of the data has been previously published and our
recalibration benefited to different degrees from improved computing
power in especially the self-calibration process.  We refer to the
original papers for any more specific information on these data:
0405$-$123, 1928+738 (Rusk 1988); 0605$-$085, 1510$-$089, and 1642+690
(O'Dea, Barvainis, \& Challis 1988); 0802+103 (Kronberg et al. 1990);
0836+299 (van Breugel et al.  1986); 2251+134 (Price et al. 1993). The
0723+679 and 1150+497 data were published by Owen
\& Puschell (1984) and were reprocessed from calibrated data obtained
directly from F. Owen. The image of 1136$-$135 was made by combining
snapshot observations published separately by Saikia et al. (1989) and
Aldcroft et al. (1993).

We were unable to locate the references for the remaining data used
and assume that they are unpublished. We list these data along with the
observer names and their \vla\ program codes: 0838+133 (J.F.C. Wardle
\& R.I.  Potash, unpublished); 1040+123 (R.L Brown: AB244); 1055+018
(W. van Breugel: AM213, B. Wills: AS396, and data from Rusk 1988),
1354+195 (P.  Barthel: AB331B, B. Wills: AS396); 1641+399
(R.A. Perley: AC120); 1741+279 (F.  Owen: AH170, and data from Price
et al. 1993).  Our 5 GHz image of 1928+738 was of poor quality so it
was supplemented with an archival 1.4 GHz \vla\ dataset (T. Rector:
AS596). The 1.4 GHz image is plotted in Figure~1. The radio flux of
the X-ray detected knot in the jet (Table 5) was obtained from the 5
GHz data and we found it to be consistent with a previous measurement
by Hummel et al. (1992).  The images presented here for 1055+018,
1136$-$135, and 1354+195 are improved versions of the data presented
in Paper I.

\section{Results}

Figure 1 shows the \chandra\ images of the 17 jets of the sample.  The
X-ray images were produced by smoothing the raw \chandra\ data with a
Gaussian of standard deviation 0.3\arcsec\ in the energy range 0.5--8
keV, with final resolution of 0.86\arcsec\ FWHM. Overlaid on the X-ray
images are the radio contours from the archival \vla\ data. Different
smoothing factors were used for individual objects; details on the
radio images are given for each source in the Appendix. 

Figure 2 shows the zoomed-in images of selected jets for which knots
close to the core were detected. The images were binned in 0.1x 0.1
pixel bins (where 1 pixel is $\sim$ 0.5\arcsec) and smoothed with
Gaussian of FWHM = 0.25\arcsec.

We define jets, knots, lobes, and hotspots on the basis of the radio
emission morphology following Bridle et al. (1994a), namely, a jet is
a narrow feature that is at least four times as long as it is wide and
a knot is a compact region of brightness in the jet. A lobe region
refers to any remaining radio emission not contained in the jet, and
is generally diffuse emission.  A hotspot is a compact feature within
the lobe. We distinguish a knot from a hotspot based on whether the
feature is located within the extended jet (a knot) or beyond its end
(a hotspot).  The position marking the end of the jet is determined by
any of the following: (1) a disappearance of emission, (2) an abrupt
change of direction (\gtsima30\deg\ within a space equal to the jet
width) for the jet emission, or (3) a decollimation of the emission by
more than a factor of two. The classification of X-ray features is
based purely on that of their presumed radio counterparts. 

As is apparent in Figure 1, X-ray counterparts to the radio knots are
detected in most sources. In other cases, only features in the lobes
are detected, either diffuse emission from the lobes themselves or
from compact hotspots within them. Details of the results for
individual sources are given in the Appendix. 

Table 2 lists the detected X-ray features in the jets. In Table 2a we
list the detected knots, while in Table 2b we list the detected
hotspots (and in a few cases, diffuse lobe emission), together with
their basic properties: name of the feature (column 3) and its
distance from the core (column 4), Position Angle (PA, column 5), and
net X-ray counts (column 6). Column 7 in Table 2 flags the features
that were also detected in the optical (Urry et al. 2004). Contrary to
Paper I, where the nomenclature was X-ray-driven, in this paper the
knots were named following the optical nomenclature. This is because
\hst\ has a higher resolution and generally detected inner knots,
which can not be directly resolved with ACIS-S. In turn, the optical
nomenclature follows previous radio publications when appropriate
(e.g., 1150+497; Akujor \& Garrington 1991). 

With respect to Paper I, the knot nomenclature changed only for
0723+679 and 1150+497. In the case of 0723+679, knot C in this paper
corresponds to knot A in Paper I, and knot D to knot B in Paper I. In
the case of 1150+497, knot B in this paper corresponds to knot A in
Paper I, knot E corresponds to knot B in Paper I, and knot H
corresponds to knot C in Paper I.

We measured the full-widths at half maximum (FWHM) of the detected
X-ray features in Table 2 in a direction perpendicular to the jet axis
(locally, as some jets bend), in the energy range 0.5--8 keV. The
widths are reported in column 8 of Table 2. For comparison, the ACIS-S
FWHM resolution is 0.5\arcsec\ at 1 keV. In most cases, the X-ray
knots are unresolved (widths \ltsima 1\arcsec), while several others
are larger than the instrumental resolution. It is worth noting that
{\it all} the knots whose radial profiles are consistent with simple
Gaussians are unresolved; the resolved knots are generally consistent
with more complex profiles, including multiple peaks (0723+679 B,
1150+497 F, 1354+195 A, B, F, 1510$-$089 B, C); most likely these
knots include multiple emission regions that are not individually
resolved with \chandra\ (but when they have an optical counterpart,
they are resolved with \hst). It is also worth noting that the knot
widths in Table 2 were derived from full-band 0.5--8 keV radial
profiles, while the encircled energy fraction decreases with
increasing energy because of larger X-ray scattering (75\% at 1.5 keV
to 65\% at 6.4 keV for an extraction radius of 1\arcsec). Therefore, a
more appropriate comparison should be performed for monochromatic
profiles; however, the signal-to-noise ratio of the present
observations is insufficient to extract radial profiles in narrow
spectral ranges. 

In the last column of Table 2a we list the photon index,
$\Gamma_{0.5-8~keV}$, from the fits to the ACIS spectra of the knots
with a power law plus Galactic N$_H$ model (see \S~3.1). Errors are
90\% confidence for one parameter of interest. The indices are
extremely flat, $\Gamma_{0.5-8~keV}$=1.3--1.6, for all knots,
indicating very hard spectra with more energy produced above the
\chandra\ band. The only exception is knot A in 1928+738, where 
$\Gamma_{0.5-8~keV}$=2.66. The average index
and 1$\sigma$ dispersions are $\langle \Gamma_{knots} \rangle =1.52$
and $\sigma_{knots}=0.21$, respectively. 

Upper limits to the X-ray counts for the radio knots and hotspots not
detected in our \chandra\ images (i.e., \ltsima $2\sigma$ detections)
are reported in Tables 3a and 3b, respectively, following the criteria
described above. Optical detections are flagged in column
8. Uncertainties are 1$\sigma$.

In all cases, the count rates in Table 2 and 3 agree with those listed
in Paper I within the uncertainties. After a careful reanalysis of the
\hst\ images, we discovered optical counterparts to the innermost
radio knots of 0723+679 and 2251+134, knots A (Urry et al. 2004). This
prompted us to report in Table 3 the corresponding X-ray counts,
determined as described in \S~3.1. In addition, for 1136$-$135 the
reanalysis of the ACIS image provided a detection in X-rays of the
innermost feature at 2.3\arcsec\ from the core, marked $\alpha$ in
Figure 1 and in Table 2a. This feature, which has a weak radio
counterpart and is undetected in the optical, is confirmed in our
deeper (80 ks) follow-up \chandra\ Cycle 4 observation (paper in
prep.).

In Paper I, it was incorrectly stated that knot I in 1354+195 (the
hotspot) was off the \hst\ field of view. In the present reanalysis we
realized that knot I falls in the \hst\ field of view, however, it is
not optically detected and none of the conclusions of Paper I are
affected.

\subsection{Detection Rates at X-rays and Optical} 

One of the goals of the survey is to establish whether high-energy
emission from kpc-scale jets is a common property given a certain
morphology and intensity of radio emission. Table~4 summarizes the
detection rates at X-rays and optical.

At X-rays, counting only the firm detections in Table 2a, the
detection rate of jets is 59\%. This number is probably a conservative
lower limit. In fact, if we add the marginal detections in Table 3a,
such as 1040+123, 1741+279, and 2251+134, where knots were tentatively
resolved in the inner regions, the detection rate of jets at X-rays
becomes 76\%.

In the optical, we find a similar detection rate (Table 4). However,
not all the knots detected at optical have X-ray counterparts, and
vice-versa. The only radio jets detected at optical, but not at
X-rays, are 1040+123 and 2251+134.  The only jets detected at X-rays,
but not at optical, are 0605--085 and 1510--089.  Only 2 sources,
0802+103 and 1055+018, have no detection at either wavelength.

In Figure 3a we plot the distribution of the radio-to-X-ray spectral
index for the jet knots, $\alpha_{rx}$, defined between 5 GHz and 1
keV (see \S~4.4 and Tables 5 and 6). The filled histograms represent
X-ray detections; upper limits are marked with arrows. As is apparent
from the Figure, the $\alpha_{rx}$ distribution is consistent with our
selection criteria ($\alpha_{rx} \sim 0.8$). Using the statistical
package \verb+asurv+ (Feigelson \& Nelson 1985), we estimated the
probability $P_{asurv}$ that the observed distribution of
$\alpha_{rx}$ in Figure 3a is drawn from a Gaussian distribution
centered on $\alpha_{rx}=0.8$, with width equal to the observed width
at half maximum in Figure 3a. Depending on the test used within the
package, $P_{asurv}$ \gtsima 90\%. The average value of $\alpha_{rx}$
is $\langle \alpha_{rx} \rangle$=0.88 with dispersion
$\sigma_{rx}$=0.08.

In Figure 3a, the non-detections span the same range of indices as for
the detections. This indicates that, due to non-uniformities in the
observations and background, even the flattest bins are not completely
covered.  Thus, there is reason to expect that essentially all bright
radio jets have associated X-ray emission, which would be apparent in
deeper and better resolved maps.

Figure 3b shows the distributions of the radio-to-optical index,
$\alpha_{ro}$, defined between 5 GHz and 5852 \AA\ (\S~4.4), for the
jet knots. As in Figure 3a, the dashed area represents solid optical
detections while the arrows indicate the upper limits to the optical
flux (Urry et al. 2004). Using \verb+asurv+, we find a probability
$P_{asurv}$ \ltsima 85\% that the observed distribution is consistent with a
Gaussian distribution centered on $\alpha_{ro}=0.8$. The average value
of $\alpha_{ro}$ in Figure 3b is $\langle \alpha_{ro}
\rangle$ = 1.05 with dispersion $\sigma_{ro}$=0.12. We conclude the
observed distribution of $\alpha_{ro}$ is consistent with steeper
values than assumed. The upper limits are distributed throughout the
full range of values for the detections. However, since fewer optical
knots were detected, determining the true distribution will require
deeper observations.


Although our sample was not optimized for X-ray/optical study of the
hotspots/lobes, we detect a fair number of them
(Table~4). Specifically, X-ray emission from hotspots/lobes was
detected in 7/17 sources, with an optical counterpart in 4
cases. Figures 3c and 3d show the distributions of the $\alpha_{rx}$
and $\alpha_{ro}$ indices for the hotspots/lobes. Both distributions
indicate steeper indices than for the jet knots; in fact, the flattest
values, easiest to detect, are missing. For Figures 3c and 3d,
$\langle \alpha_{rx} \rangle$ = 1.03 and $\sigma_{rx}$=0.05 and $\langle
\alpha_{ro} \rangle$ = 1.17 and $\sigma_{rx}$=0.15. 

Finally, in 5/17 sources there is a weak X-ray detection of the
counterlobe (0723+679, 0838+133, 0836+299, 1040+123, and
1136--135). X-ray emission from the counterlobe of 0838+133 was
interpreted by Brunetti et al. (2002) as back-scattered Compton flux,
and a similar explanation may hold for 0723+679 (Paper I). While
0723+679, 0838+133, 1040+123, and 1136--135 are powerful FRIIs, the
radio galaxy 0836+299 exhibits an FRI morphology and has a radio power
similar to an FRI (van Breugel et al. 1986). Both hotspots in the lobe
and counterlobe in 0836+299 have an optical counterpart.  However, the
optical emission from the Northern hotspot is probably due to emission
lines (van Breugel et al. 1986). Since our very broad filter does not
distinguish emission lines from continuum, we do not quote an optical
flux for this feature.

\subsection{Jet Multiwavelength Morphologies} 

\noindent{\bf X-rays versus radio:} Concentrating on a comparison of
the radio and X-ray jets in Figure~1, a variety of morphologies is 
apparent. In most cases, the X-rays track the radio one-to-one, i.e.,
all X-ray knots have a radio counterpart. 

The jet of 1136$-$135 stands out for its remarkable multiwavelength
morphology. As clearly shown in Figure 1, there is weak or no radio
emission from the inner jet which instead is detected at X-rays (knots
$\alpha$, A), and while after knot B at $\sim$ 7\arcsec\ from the core
the X-rays start to fade, the radio emission picks up, peaking at
$\sim$ 10\arcsec\ from the nucleus. A similar situation occurs for
1510$-$089, where the jet morphology is rather similar at radio and
X-rays in the inner parts while the radio-to-X-ray flux density ratio
increases dramatically after 8\arcsec. 
3C~273 is another example of this morphology (Sambruna et al. 2001).

In 1040+123, 1741+279, and 2251+134, the X-ray knots are not resolved
as they are located at \ltsima 1\arcsec\ from the strong core, and we
can not comment on these jets morphology. Higher-resolution X-ray
observations are needed to study these jets.

\noindent{\bf X-rays versus optical:} As discussed above, seven jets
have detections at both X-rays and optical, while two have only 
optical counterparts. 
For most of the optical/X-ray jets, optical emission is confined
within 3--4\arcsec\ from the nucleus, while X-ray emission extends to
larger distances. This may be due to the broader $\alpha_{ro}$
distribution where a detection would require a deeper \hst\
observation. Because of the many non-detections we cannot distinguish
between a steep $\alpha_{ro}$ (too steep for the short \hst\ exposure)
and a true lack of optical emission from the jet. The three exceptions
are 1040+123, 1136$-$135, and 1150+497, where the optical jet is as
long as (1150+497) or longer then (1040+123, 1136$-$135) the X-ray
jet.

Recent \chandra\ studies showed that X-ray emission from FRI jets may
be common (e.g., Worrall et al. 2001). In the only FRI source of our
sample, 0836+299, weak X-ray excess counts over the background are
detected in the ACIS image in correspondence to the inner radio jet at
$\sim$ 2\arcsec\ (Figure 1), well inside the optical galaxy. However,
a similar X-ray feature is also present at opposite azimuth, raising
the possibility that the X-ray emission from the radio feature is
spurious. No optical counterpart is detected in our \hst\ image, after
subtracting the host galaxy.

\subsection{Continuous Intra-knot X-ray Emission} 

Interestingly, in a few cases there is continuous, weak intra-knot
X-ray emission. A clear example is 0605$-$085, where the inner jet
between knots A and B shows smooth X-ray emission, with no apparent
compact knots. Other candidates for continuous intra-knot X-ray
emission are 0838+133 and 1136$-$135; however, in 1136$-$135 the
optical knots are compact. While the limited ACIS resolution may
conspire to produce diffuse emission where many small compact knots
are instead present (or smearing out the photons from nearby strong
knots), in 0605$-$085\ the extension of the continuous X-ray emission
is at least 2\arcsec, larger than the S3 FWHM resolution, and so it is
significant in this source. A brief discussion of the possible origin of
the intraknot X-ray emission in given below (\S~5.1). 

Emission from the inner jet was also detected in PKS 0637$-$752 in a
100-ks ACIS-S exposure (Chartas et al. 2000), with a different
spectrum than the external part of the jet, and in 3C~273 in the inner
10\arcsec\ (Marshall et al. 2001). In the latter object, the inner
radio/X-ray jet was recently detected in the optical band with the ACS
camera on \hst\ (Martel et al. 2003).

\subsection{Spectral Energy Distributions} 

The Spectral Energy Distributions (SEDs) of the radiation emitted by
the detected features provide basic information on the emission
mechanisms. We caution, however, that the derived flux densities refer
mainly to unresolved sources, thus the size of the emission region is
uncertain, at least at X-rays. In the optical, most of the knots are
unresolved even at the \hst\ resolution (0.2\arcsec; Urry et
al. 2004), and we will use an upper limit to the source size in the
modeling (see below). For consistency, we extracted flux densities at
the three wavelengths from the same spatial region around the
knot. Since the X-rays have the lowest resolution, the aperture of the
extraction region was fixed at 1\arcsec\ or 1.5\arcsec. The extraction
region was centered on the position of the X-ray knot, or of the
optical one when no X-ray feature was detected.

Following Paper I, radio fluxes for those features well separated from
the nucleus were extracted from the same aperture as the X-ray flux
(see section 3.1). The optical fluxes were extracted as described in
\S~3.2. The unabsorbed X-ray flux at 1 keV 
was derived from the fit to the ACIS spectrum of the knot, when
available (Table 2a), or using a power law with average X-ray photon
index, $\langle \Gamma_{knots} \rangle =1.52$. The optical fluxes are
corrected for Galactic extinction. Intrinsic reddening, impossible to
estimate with the data in hand, is highly unlikely as the line of
sight is close to the jet axis and the resolved knots are at large
distances from the nucleus.

The optical flux plays a critical role in the interpretation of the
SED.  Specifically, if the optical emission lies on the extrapolation
between the radio and X-ray fluxes or above it, the SED is compatible
with a single electron spectrum extending to high energies; instead,
if the optical emission falls well below the extrapolation, it argues
for different spectral components (and therefore different mechanisms
or two electron populations) below and above the optical range (e.g.,
synchrotron and IC respectively). Thus, an up-turn of the spectrum in
the X-ray band with respect to the radio-optical extrapolation,
yielding an optical-to-X-ray index $\alpha_{ox}$ flatter than the
radio-to-optical index $\alpha_{ro}$, is a signature of a separate
spectral component in the X-ray band. Conversely, when synchrotron
dominates we expect $\alpha_{ro}$ \simlt $\alpha_{ox}$, with the
inequality holding when radiative losses are important in the X-ray
band.

Table 5a-b lists the radio, optical, and X-ray flux densities, or upper
limits to them, for the jet knots, while Table 6a-b lists the same for
the lobes. Also listed in both Tables are the broad-band indices
$\alpha_{ro}$, $\alpha_{ox}$, and $\alpha_{rx}$, defined as the
spectral indices between 5 GHz and 5852 \AA, 5852 \AA\ and 1 keV, and
5 GHz and 1 keV, respectively.

Figure~4 shows the plot of $\alpha_{ro}$ versus $\alpha_{ox}$ for all
the jet features for which a firm detection at either optical or
X-rays is available. 
The dotted line, corresponding to $\alpha_{ro}=\alpha_{ox}$, separates
the SEDs of the various features between concave (left) and convex
(right). In the large majority of the jet knots, X-rays are not an
extension of the radio-to-optical synchrotron spectrum; the X-ray flux
lies above the extrapolation from the radio-to-optical
slope. Similarly, most hotspots have concave SEDs.

Figure~5 shows the relative variation of the X-ray-to-radio flux
ratios along the jet. In the Figure, $\alpha_{rx}$ is plotted vs. the
projected distance of the knots from the core. The uncertainties on
$\alpha_{rx}$ are listed in Table 5. Remarkably, a trend is present in
each jet of {\it decreasing X-ray-to-radio flux (increasing
$\alpha_{rx}$) from the innermost to the outermost regions of the
jet}. Note that the trend is present at large distances from the core,
where contamination of the X-ray flux from the PSF wings is
negligible. A similar but less robust trend of decreasing
optical-to-radio flux appears in Figure~6, where $\alpha_{ro}$ is
plotted versus the projected distance. We will comment on these trends
in \S~5.3.


\section{Origin of the X-ray emission}

\subsection{Clues to the X-ray emission mechanism} 

As seen in Figure 1, the detected jets exhibit a variety of
multiwavelength morphologies. In a sense, each jet appears to be
unique in its detailed properties and deserves a detailed individual
study. In fact, we have already secured deeper follow-up observations
in \chandra\ AO4 and multi-color \hst\ ACS exposures of 1136$-$135 and
1150+497.

General clues to the origin of the X-ray emission can be offered by
the study of multiwavelength SEDs for the different emission features
along the jet.  It is well established that the radio emission is due
to the synchrotron mechanism.  Thus, the simplest hypothesis to be
considered is whether the optical and X-ray emission could be due to
synchrotron emission from the same electron distribution (in a simple,
zero-order approximation).  In that case one would expect the radiated
spectrum to follow a power law or to steepen at higher energies due to
radiation losses.  From the plot of $\alpha_{ro}$ versus $\alpha_{ox}$
(Figure~4) it appears that this may be true for at most two cases
which fall close to the line for which $\alpha_{ro}=\alpha_{ox}$. One
of them corresponds to knot A in the jet of 1136$-$135, the other to
knot A of 1928+738. However, the vast majority of knots lie in the
region $\alpha_{ro} > \alpha_{ox}$, showing X-ray emission {\it in
excess} of the extrapolation from lower energies.

In order to explain the excess X-ray radiation via the synchrotron
mechanism, it is necessary to invoke an extra component in the
population of relativistic electrons (e.g., Wilson \& Yang 2002). A
particularly elegant possibility was proposed by Dermer \& Atoyan
(2002), whereby relativistic electrons are continuously injected and
subject to radiative losses dominated by inverse-Compton
cooling. However, due to the Klein-Nishina suppression of the Compton
cooling rate, high-energy electrons suffer less cooling compared to
low-energy ones, naturally developing a high-energy excess in the
spectrum. In this hypothesis the observed X-rays would be produced by
very-high energy electrons.

Alternatively, in a synchrotron plus inverse Compton (IC) scenario, a
single power-law electron population can be responsible for emitting
the radio via synchrotron and the X-rays via IC, any external seed
photon field being amplified if the jet is still relativistic on the
relevant scale. Note that these knots occur predominantly at large
distances from the core. As the jets are very long (projected lengths
$\sim$ 50--100 kpc, intrinsic lengths possibly 2-3 times longer,
perhaps 10 times longer) and extend outside the host galaxy, the most
likely source of seed photons for IC is provided by the Cosmic
Microwave Background (CMB) photons (Tavecchio et al. 2000), whose
energy density scales as $(1+z)^4$. For plausible values of the
Doppler factor and the magnetic field ($\delta \sim 10$, $B\sim 10
\mu$G), the energy of the electrons radiating via synchrotron in the
radio and via IC in the X-rays is expected to be relatively close
($\gamma _r=1000$, $\gamma _X=100$), hence a similar jet morphology at
both wavelengths is natural, as observed in most knots. X-rays from
these features are therefore likely explained by the IC/CMB model, on
both spectral and morphological grounds.
However, starlight photons can be important for knots within the
galaxy, within a few tens of kpc of the core (estimated in a similar
way to Stawarz, Sikora, \& Ostrowski 2003). These photons would be
upscattered in the X-ray band by very low-energy electrons, $\gamma
\sim 2$. Due to the limited ACIS resolution, emission from the
innermost knots cannot be easily quantified.

A discriminant between synchrotron and IC origin for the X-ray
emission is the particle radiative lifetime. One would expect shorter
radiative lifetimes, and thus more compact emission regions, at the
shorter wavelengths in the synchrotron model, whereby X-rays derive
from high-energy electrons. The morphology of the 1136--135 jet
suggests synchrotron is important in at least some knots. This is
supported by the radio-to-X-ray spectral energy distribution of knot A
in 1136$-$135, which indeed does not show an X-ray excess and was
fitted by synchrotron emission in Paper I.

A direct consequence of the IC/CMB model is that X-rays are produced
by low-energy electrons, characterized by an extremely large cooling
timescale. Since the electrons can not cool within the short distance
of a knot, we expect continuous X-ray emission due to their streaming
along the jet.  Indeed, in a few cases intraknot emission is visible
in our \chandra\ images (\S~4.3). Therefore, it is possible to
accommodate the knotty morphology with the virtually infinite cooling
time of X-ray electrons, assuming that knots just represent local
enhancement of the surface brightness, as expected in a shock
compression scenario. Due to the limited sensitivity of our data, we
can not draw firm conclusion.  As discussed in Tavecchio, Ghisellini,
\& Celotti (2003), the presence of isolated knots would represent a
severe problem for the simplest version of the IC/CMB model. A
possible way out would be to postulate that the low-energy electrons
are cooled through adiabatic losses, effective only if the source can
expand enough to cool electrons from $\gamma
\sim 100$ to $\gamma \sim 10$. These considerations lead Tavecchio et
al. (2003) to propose that the emission from a ``knot'' is instead due
to a large number of unresolved expanding ``clumps''. Deeper
observations at higher resolution are needed to further investigate
possible knot substructures. 

\subsection{Reproducing the SEDs} 

To investigate quantitatively the jet physical properties within the
simple homogeneous synchrotron + IC/CMB model, we modeled the SEDs of
the knots which were detected at both optical and X-rays. Clearly,
with only three measured fluxes the models are underconstrained.
However, in the cases where X-rays can be attributed to IC/CMB the
model parameters can be fixed if the equipartition assumption is
adopted.

A significant uncertainty of present observations concerns the size of
the emitting region(s). In what follows, we assume that the emitting
region is consistent at the three wavelengths examined and adopt a
region size of 1\arcsec\ radius, corresponding to the X-ray resolution
and the area from which fluxes were extracted.

We refer to Tavecchio et al. (2000) and Paper I for a full description
of the model. Briefly, the emitting region is assumed to be spherical
with radius $R$ and in motion with a bulk Lorentz factor $\Gamma$ at
an angle $\theta$ with respect to the line of sight. Since fluxes are
extracted within a circle of radius 1\arcsec, we fix the dimension $R$
(in cm) corresponding to this angular size. Note that this is
different than Paper I, where for simplicity we considered a unique
value of radius for all the sources. The emitting region is
homogeneously filled by high-energy electrons, with a power-law energy
distribution $N(\gamma )=K\gamma^{-n}$ extending from $\gamma_{\rm
min}$ to $\gamma_{\rm max}$. Electrons emit radiation through
synchrotron and IC/CMB mechanisms.  The low-energy limit of the
electron distribution $\gamma_{\rm min}$ is well constrained by the
condition that the low-energy part of the IC/CMB component cannot
overproduce the observed optical flux. Note that the optical emission
in many cases could be attributed either to the high-energy tail of
the synchrotron component (in this case the optical spectrum is
expected to be soft) or to the low-energy tail of the IC/CMB component
(hard optical spectrum expected). The X-ray spectra of the knots in
Table 2a are relatively flat in all the cases where the SEDs are
consistent with IC/CMB, as expected (e.g., Fig. 8 in Paper I). The
parameters of the best-fit models are reported in Table 7. Figure 7
shows representative SEDs for two sources where at least two knots
were detected at X-rays and optical in the same jet, with the best-fit
models superposed. 

For two cases, namely knots A in 1136$-$135 and 1928+738, the
optical-radio-X-ray fluxes are consistent with one component,
indicating the X-rays are produced via synchrotron. Therefore, in
these cases we have reproduced the observed fluxes imposing that the
IC/CMB component does not substantially contribute to the X-ray flux:
This provides a lower limit on the Doppler factor, that translates
into a limit on the bulk Lorentz factor and on the observing angle
reported in Table 7.  The kinetic and radiative powers derived with
the parameters reported in Table 7 confirm the evidence (see also
Paper I) that only a small fraction (\ltsima 0.1\%) of the jet kinetic
power is dissipated into radiation.


Since our choice of the size of the emitting region based on the
\chandra\ resolution is rather restrictive, we checked the sensitivity
of the derived parameters on the assumed volume. Since we know from the
optical images that the (optical) knots are unresolved down to $\sim
0.2$\arcsec, we performed a series of fits assuming a cylindrical
region with height 2\arcsec\ (along the jet direction) and radius
0.2\arcsec. The results of this new set of fits show that the derived
parameters are not strongly affected by this change in volume: in
particular $\delta$ is slighly larger (about a factor 1.5) and
similarly the magnetic field and the electron density change within a
factor of 3--4.


\subsection{Trends of the emission along the jet}

If emission from single knots seems to be well reproduced by the
IC/CMB model, further interesting clues to the origin of the
high-energy emission and to the global dynamics of the jet originate
from examination the overall trends of the SEDs for different knots
along the same jet. These were presented in Figures 5 and 6, which
show the plots of $\alpha_{rx}$ and $\alpha_{ro}$ versus the projected
distance of the knots from the cores. As discussed in \S~4.4, the
X-ray-to-radio and optical-to-radio flux ratio decreases along the
jets.

The trends in the two plots suggest that a common evolution of the
conditions along the jet determines the observed properties of the
emission features.  In a pure ``synchrotron+IC/CMB'' model, the
relative strength of the X-ray and radio emission, which to first
order is given by the ratio of the CMB and magnetic densities, is
expected to increase, since the magnetic field within the jet
presumably decreases and the energy density of the CMB is constant.
The observed {\it opposite} behaviors could have different origins:

\begin{itemize} 

\item It could be due to the synchrotron mechanism extending to the
X-ray band in the inner knots (due to a higher value of the magnetic
field and/or the particle energies), thus providing an additional
X-ray component which progressively disappears in the more external
knots.  This scenario is supported by the case of 1136--135, where a
synchrotron emitting inner knot is resolved with ACIS. This
interpretation is also supported by the case of 3C~273, where the
X-ray emission from the first knot is possibly synchrotron (Marshall
et al. 2001). The preliminary analysis of our deeper \chandra\ and
multicolor \hst\ exposures of 1150+497 and 1136--135 also confirm this
interpretation.

\item An additional component of IC emission could be provided in 
the innermost knots by the interstellar light photons whose energy
density within the galaxy can prevail over the CMB density
(e.g. Stawarz et al. 2003).  Along the jet this contribution will
decrease, producing the observed trend of the X-ray-to-radio flux
ratio.

\item In a pure IC/CMB scenario, the trend could be attributed to
decelerating plasma: the decreasing value of $\Gamma$ translates into
a less amplified CMB radiation and thus a lower X-ray flux compared to
the radio flux.  Elements supporting this view are provided by our
modeling of the jet of 1354+195, for which two distinct knots were
analyzed (see Table~7). The data require that the bulk Lorentz factor
between the two features decrease by a factor of 2, from $\Gamma =14$
(knot A) to $\Gamma =6$ (knot B). At the same time the magnetic field
strength and the number of particles {\it increase}. (However,
remember that we are assuming equipartition, so the magnetic energy
density and the electron energy density are linked.)

\end{itemize} 

The trend observed in the $\alpha_{ro}$ profiles can simply be related
to the decreasing value of the maximum Lorentz factor of the emitting
electrons. This will produce a shift of the synchrotron cutoff toward
lower frequencies along the jet and therefore a decreasing optical
flux along the jet (note that the same beavior can be mimicked by a
decrease of the magnetic field). A better understanding of these and
other systematic trends requires more sophisticated theoretical
studies than is possible to include here.

\subsection{Caveats}

It is worth remarking a few caveats affecting our analysis.  First,
the limited signal-to-noise ratio at both X-ray and optical
wavelengths leaves room for alternative interpretations of the
SEDs. Second, as already mentioned above, a variety of physical
conditions may exist within the relatively large extraction regions we
used (1\arcsec, dictated by the \chandra\ resolution), for example the
emitting particle distributions could be stratified or multiple shocks
may exist. While higher angular resolution at X-rays awaits future
generations of space-based telescopes, deeper follow-up X-ray and
optical observations of the new jets of this survey with \chandra\ and
\hst\ can at least remedy the first limitation of our analysis, in
providing accurate X-ray and optical continuum spectra for individual
knots, a key test for the emission models.

High-quality X-ray and optical spectra of single knots will be
essential to discriminate among the various models, as well as more
detailed maps to measure and quantify the positional offsets of the
radio, optical, and X-ray peaks. Optical observations are necessary to
identify the mechanism responsible for the X-ray emission, as the
optical band lies at the intersection of the synchrotron and IC
components. We have already taken steps to acquire follow-up
observations of selected jets during the Cycle 4 \chandra-\hst\
cycle. Finally, an additional important constraint will be provided by
future IR observations with \sirtf, probing a poorly known region in
the SEDs where the synchrotron peak (related to the break energy of
the synchrotron electron population) is located. Of particular
interest will be \sirtf\ observations of the jets {\it not} detected
at optical wavelengths, to determine whether this can be due to a
lower cutoff energy of the electron distribution.

\subsection{Comparison with FRI Jets}

\chandra\ has detected X-ray emission in a handful of low-power FRI
sources (Worrall et al. 2001, Hardcastle et al. 2002, 2001). It is
interesting to compare their properties to the FRII jets of our sample.

In most FRI jets, the X-ray flux of the detected knots lies on the
extrapolation of the radio-to-optical emission.  The favored
interpretation is that the X-rays are due to synchrotron emission from
the same population of electrons responsible for the radio and optical
fluxes. The X-ray spectrum of the knot, when available, is also
consistent with the low-energy emission, confirming this
interpretation. The radio/X-ray morphology of the jet is remarkably
similar in all FRIs. In all cases, the X-ray profile of the first
detected knot peaks before the radio, yielding a larger X-ray-to-radio
flux ratio, or flatter $\alpha_{rx}$, than in the knots further away
along the jet.  A similar situation occurs for sources close to the
FRI/II transition (e.g., 3C~371, Pesce et al. 2001; PKS~0521--365,
Birkinshaw et al. 2002).

Thus, the X-ray emission from kpc-scale jets in FRIs and FRIIs appears
to be due to different processes, with synchrotron dominating in
low-power sources and IC/CMB in high-power sources.
However, a closer look to our targets reveals that the FRI/II division
of jet emitting mechanisms may not be sharp. Indeed, as discussed
above, the inner parts of the powerful FRII jets may be dominated by
different production mechanisms for the X-rays, suggesting a change of
conditions along the jet. A clear example from our survey is
1136$-$135. In this case, X-rays from the innermost knot A ($\sim$
5\arcsec, or 25 kpc) are due to synchrotron emission of high-energy
electrons in the jet, while X-rays from external knot B are consistent
with IC/CMB (Table 7).  Strong particle acceleration is necessary to
explain synchrotron X-rays in the inner parts of the jet.

A possibility to account for the FRI/II dichotomy on large-scales is
proposed in our companion paper (Tavecchio et al. 2004). We suggest
that the main difference between FRI and FRII jets is the location at
which the jet pressure becomes comparable to the ambient gas pressure
and the jet starts to slow down significantly, giving rise to
shocks. Assuming that both FRI and FRII host galaxies have similar gas
halos (e.g., Gambill et al. 2003), the discriminating parameter
becomes the jet pressure which is related to the jet power (Maraschi
\& Tavecchio 2003): the more luminous sources (FRIIs) are slowed down
later than less luminous ones. If this is true, one expects to observe
the sites where dissipation is taking place - the innermost knots -
closer to the core in lower-luminosity sources than in the more
powerful ones.

To illustrate this point, we plot in Figure 8 the total radio power of
the sources versus the {\it deprojected} distance of the first
detected knot (optical or X-ray). The total radio powers were derived
from Cheung et al. (2004), Liu \& Xie (1992), and Liu \& Zhang
(2002). Only the sources for which an estimate of the angle from
modeling is available (Table 7) were used. In addition, we plot the
same quantities for a few low-power radio galaxies from the literature
for which enough information is available (3C~371, Pesce et al. 2001;
M87, Wilson \& Yang 2002; 3C~31, 3C~66B, Hardcastle et al. 2002, 2001;
and NGC~315, Worrall et al. 2003). The only FRI of our sample,
0836+299, is also plotted (triangle). As apparent from Figure 8, low-
and high-power sources occupy distinct regions, with the FRIIs having
the first detected knots at larger distances from the nucleus than
lower-luminosity sources. A trend is also present within the FRII
class itself.

We note, however, that the FRIIs in the Figure are at larger redshifts
than FRIs. The loss in resolution thus could be introducing a bias,
i.e., we are not resolving the innermost knots in the most distant
FRIIs. In fact, in 3C~273 ($z$=0.158) optical knots very close
(2.7\arcsec, or projected distance of 7 kpc) to the nucleus were
recently detected with the ACS camera on \hst\ (Martel et
al. 2003). Clearly, high-resolution sensitive observations of the
inner jets in FRIIs are needed to confirm the suggestion of Figure
8. Moreover, the deprojected jet length clearly depends on the assumed
(hence model-dependent) value of the viewing angle. In this respect,
the choice of the IC/CMB model for the high-power datapoints in Figure
8 could introduce a bias toward small angles and therefore large
deprojected lengths. However, other (almost model-independent)
indicators such as the core dominance parameter $R_i$ and the
jet-to-counter jet ratio $J$ (Table 1), suggest small observing
angles. Based on these considerations, we are confident that the large
deprojected lengths in Figure~8, albeit affected by inevitable
uncertainties, are reliable.

Although the scenario depicted above can explain some of the basic
differences between FRIs and FRII, the study of the nearest FRI, M87,
shows that the situation may be more complex. In this object the X-ray
flux belongs to a separate component than the longer wavelength
flux. The steep X-ray spectrum suggests synchrotron emission from a
separate population of particles (Wilson \& Yang 2002), indicating jet
inhomogeneities. Moreover, we recall that there are also indications
that the jet can have a velocity structure, with a fast spine
surrounded by a slower moving layer (as originally proposed by Laing
et al. 1993). Synchrotron emission would be produced in the walls of
the jet, while IC/CMB would predominate in the beamed emission
concentrated in the spine.

\section{Summary and Conclusions}

We presented short \chandra\ and \hst\ observations of a sample of 17
radio jets. One goal of the survey was to establish in a systematic
way the relation between X-ray/optical and radio emission from
extragalactic jets.  The survey was therefore designed to provide at
least a detection in X-rays for a radio-to-X-ray spectral index
$\alpha_{rx}=0.8$, and in the optical for $\alpha_{ro}=0.8$.  The
short exposures were designed to search for the X-ray and optical
counterparts of the radio jets, and start addressing their physical
properties through multiwavelength imaging.

We detect X-ray emitting knots in 10/17 (59\%) jets, with a similar
detection rate for the optical features. This is actually a lower
limit to the number of X-ray emitting knots, as several radio- and
optically-detected features are too close to the core (\ltsima
1--2\arcsec) to be clearly resolved with ACIS. Moreover, as discussed
above, the non-detections are most likely related to incompleteness
rather than $\alpha_{rx}$ indices steeper than expected (Figures
3a). We can thus conclude that {\it X-ray emission from extended radio
jets with $\alpha_{rx} \sim 0.8$ is common in AGN}, and it is consistent
with being a characteristic of essentially all bright radio jets.

The rate of detection in Table 2a is similar for FSRQs and SSRQs,
$\sim$ 60\%. The presence of a large number of FSRQs in the sample
(Table 1) is a direct indication that our sample is biased towards
beamed objects. Thus, the detection rates for jets refer to lines of
sight close to the jet axis at both X-rays and optical. This is in
agreement with previous optical studies (Parma et al. 2003, Sparks et
al. 1995), which concluded that jet optical emission is preferentially
detected in those sources with larger beaming, i.e., where the jet is
closer to the line of sight. In fact, the two jets in our sample with
the lowest values of the core-to-total radio power, 0802+103
($R_i \sim 0.03$) and 0836+299 ($R_i \sim 0.06$), are not detected at
either optical or X-rays.

In our sample, there is no obvious dependence of the detection rates
on redshift. The most distant jet (0802+103 at $z$=1.96) is not
detected at X-rays or optical, however, this source has a lesser
defined radio jet (Figure 1) and is the most lobe-dominated source in
the sample. The second most distant, 1040+123 at $z$=1.03, is
well-detected at optical wavelengths but only upper limits are present
at X-rays. The third most distant, 1055+018 at $z$=0.88, is not
detected. A larger, systematic sample of high-redshift sources with
well-defined radio data is needed to confirm the suggestion (Schwartz
2002) that X-ray emission is common at cosmological redshifts as a
result of an increased CMB density.


In summary, the main results of our paper are: 

\begin{itemize} 

\item X-ray and optical emission is detected in $\sim$ 60\% of the jets, with
$\langle \alpha_{ro} \rangle \sim 1.0$ and $\langle \alpha_{rx}
\rangle \sim 0.8$.  

\item The non-detections in the X-rays 
are due to incompleteness rather than $\alpha_{rx}$ indices steeper
than expected, meaning essentially all radio jets have X-ray emission
at some intensity level. The observed distribution of $\alpha_{ro}$
indices is consistent with steeper values than assumed.

\item In most knots the X-ray flux lies above the extrapolation of
the radio-to-optical continuum, indicating a separate spectral
component. 

\item Interpreting the X-ray emission of these knots as due to
inverse Compton scattering off the CMB photons, yields relativistic
jets on kpc scales, with plasma Lorentz factors $\Gamma_{jet} \sim
3-15$, and very low ($<$ 0.1\%) radiative efficiencies.

\item In a few knots the X-ray flux lies below or on the extrapolation of
the radio-to-optical continuum. Here the X-rays are likely to be due
to synchrotron emission from the same population of relativistic
particles responsible for the longer wavelengths.

\item Spectral changes along the jets are observed, with the
X-ray-to-radio and optical-to-radio flux ratios decreasing from the
inner to the outer regions. 

\item The trend of decreasing X-ray-to-radio flux ratios can be
either the results of 1) an additional X-ray producing process
in the inner jet (synchrotron or inverse Compton scattering of
starlight), or 2) in the IC/CMB scenario, decelerating plasma. The
trend in $\alpha_{ro}$ could simply be due to a decrease of the
electrons Lorentz factor along the jet. 


\end{itemize} 

Several questions remain open. Most fundamental one is the origin of
the bright X-ray emission from kpc-scale jets. As discussed above, the
IC/CMB process appears to account satisfactorily for the flat
optical-to-X-ray spectral indices of most knots, at least in the
external part of the jets, but implies the presence of relativistic
bulk motion at large distances from the core. This is in contrast with
radio studies showing only modest bulk velocities at such distances
(e.g., Wardle \& Aaron 1997), but is consistent with jet one-sidedness
and with depolarization asymmetries (Garrington et
al. 1988). Additional contributions to the X-ray emission may be
present along the jet, e.g., IC of the host galaxy starlight, which
could account for the decreasing X-ray-to-radio flux ratios along the
jet (see above). Possible inhomogeneities in the jet (multiple shocks
or particle distributions) in space or/and time are additional
complicating factors.

In some FRII jets we observe ``mixed'' emission with synchrotron
dominating in the inner parts of the jet (e.g., 1136--135,
3C~273). The rate of occurrence of this is still unclear, as is its
physical origin. 


Finally, we note that the X-ray spectral indices measured in the large
majority of the jets knots are flat, $\alpha_X \sim 0.5$, indicating
that most of the jet luminosity is emitted at $\gamma$-rays. This
raises the interesting possibility of future GLAST detections. Given
that radio jets might be bright at large redshifts (Schwartz 2002), it
would not be surprising if extended jets would be found to be
significant contributors to the $\gamma$-ray background. While GLAST
lacks sufficient angular resolution to separate the kpc-scale jet from
the bright core, variability studies could conceivably be used to
discriminate the origin of the gamma-ray emission.


\acknowledgements 

The referee, Dan Harris, provided constructive criticism which helped
improving the manuscript.  We thank M. Gliozzi for help in running the
\verb+asurv+ package and in the revision stage of the paper. This
project is funded by NASA grant HST-GO-12110A, which is operated by
AURA, Inc., under NASA contract NAS 5-26555, and by grant NAS8-39073
(RMS and JKG). RMS gratefully acknowledges support from an NSF CAREER
award and from the Clare Boothe Luce Program of the Henry Luce
Foundation.  Radio astronomy at Brandeis University is funded by the
NSF. Funds by NASA grants GO2-3195C from the Smithsonian Observatory
(CCC), NAG5-9327 (CMU), and HST-GO-09122.01-A (CMU, CCC) are
gratefully acknowledged.  IRAF is distributed by the National Optical
Astronomy Observatories, which are operated by the Association of
Universities for Research in Astronomy, Inc., under cooperative
agreement with the National Science Foundation.

\noindent{\bf Appendix: Comments to Individual Sources} 

\noindent {\bf 0405$-$123:} The \vla\ contours in Figure~1 were
restored with beamsize 1.25\arcsec. The lowest contour is 0.5
mJy/beam. Only the hotspot of the northern lobe was detected at X-rays and
optical. The X-ray/optical feature coincides well with the radio one,
with no centroid offset.

\noindent {\bf 0605$-$085:} The \vla\ contours in Figure~1 were
restored with beamsize 0.45\arcsec. The lowest contour is 0.75
mJy/beam. Two X-ray knots were detected, with no optical counterpart
in the \hst\ image, as well as faint intraknot emission. The X-ray
emission peaks 4.0\arcsec(B), before the radio knot which peaks at
4.5\arcsec.  This radio knot is centered with a difference of position
angle, $\Delta$PA=18\deg, further south.  The nearby bright source on
the S-W is a foreground star, with a bright optical counterpart on the
\hst\ map.

\noindent {\bf 0723+679:} The \vla\ contours in Figure~1 were
restored with beamsize 0.5\arcsec. The lowest contour is 0.09
mJy/beam. The radio images were reprocessed from calibrated data
published in Owen \& Puschell (1984). Knots C and D at 4.4\arcsec\ and
6.6\arcsec, respectively, were already reported in Paper I (as A and
B). Analysis of the \hst\ image, showing an optical knot at
0.9\arcsec\ (knot A), prompted a reanalysis of the inner jet in the
ACIS image. Following the method described above, knot A is detected
at X-rays, although it is not resolved. As contamination from the PSF
wings could affect out measurement, we list the measured counts as an
upper limit (Table 3a).  An additional knot is detected at 2\arcsec\
from the core (knot B in Table 2). Weak, diffuse X-ray emission from
the counterlobe is also present (Table 3).

\noindent {\bf 0802+103:} The \vla\ contours in Figure~1 were
restored with beamsize 0.3\arcsec. The lowest contour is 0.2
mJy/beam. Weak, fuzzy X-ray emission, generally elongated in the same
direction as the South radio jet, is present in Figure 1. The enhanced
X-ray counts correspond to the edge of the jet. It is unclear whether
this X-ray counts are associated to the jet or represent a background
fluctuation or background source. Some diffuse X-ray emission is also
present in the N counterlobe, again with no clear correspondence to
the radio. We report upper limits to both X-ray structures in Table 3.

\noindent {\bf 0836+299:} The \vla\ contours in Figure~1 were
restored with beamsize 0.86\arcsec. The lowest contour is 0.08
mJy/beam. \chandra\ clearly detects X-ray emission from the hotspots
in the S-W radio lobe at 17.8\arcsec, and in the counterlobe at
11.3\arcsec. Both hotspots have an optical detection in our \hst\
data. The ACIS map shows a faint X-ray structure (A) generally
coincident with the inner portion of the S-W radio jet (Figure 1),
with less than $2\sigma$ significance (Table 3). However, the presence
of a nearly identical (same number of counts) feature in the opposite
direction makes the X-ray emission from the jet uncertain.

\noindent {\bf 0838+133:} The \vla\ contours in Figure~1 were
restored with beamsize 0.35\arcsec. The lowest contour is 0.3
mJy/beam. X-ray counterparts to three radio knots at 1.4\arcsec,
4.6\arcsec, and 6.5\arcsec\ are detected in the \chandra\ image. Knot
A at 1.4\arcsec\ also emits in the optical.  Feature C is identified
as a hotspot. Weak emission from the counterlobe is also confirmed
(Brunetti et al. 2002).

\noindent {\bf 1040+123:} The \vla\ contours in Figure~1 were
restored with beamsize 0.25\arcsec. The lowest contour is 0.65
mJy/beam. The \hst\ data show optical emission from several radio
knots at 0.9\arcsec\ (B), 1.5\arcsec\ (C), and 4.8\arcsec\ (D). In the
\chandra\ image (Figure 1), knot B is marginally detected, 
but not resolved given its proximity to the strong X-ray core. Fuzzy
X-ray emission is visible around the counterlobe, although it may be a
background fluctuation or foreground sources.

\noindent {\bf 1055+018:} The \vla\ contours in Figure~1 were
restored with beamsize 1.5\arcsec. The lowest contour is 0.85
mJy/beam. This source was previously studied in Paper I. No X-ray or
optical jet is detected (Table 3a).  Paper I lists an upper limit to the jet.

\noindent {\bf 1136$-$135:} The \vla\ contours in Figure~1 were
restored with beamsize 0.5\arcsec. The lowest contour is 0.5
mJy/beam. We confirm the results of Paper I. X-ray emission is
detected from two knots at 4.6\arcsec\ (A) and 6.7\arcsec\ (B) from
the core, both of which have an optical counterpart.  A third knot (C)
is detected in the optical at the end of the radio jet, at 10\arcsec;
however, no X-ray counterpart is present. Our reanalysis of the
\chandra\ image also showed the presence of an inner knot, $\alpha$
(Table 2), which is confirmed by our deep (80 ks) GO4 exposure.  In
the radio, knot $\alpha$ has a 4$\sigma$ detection in our 5 GHz \vla\
image at 0.4 mJy, though it is not obvious in Figure 1. It was
detected similarly in our unpublished deep 22 GHz \vla\ image. The jet
X-ray emission peaks at knot B while the radio peaks at knot C, at the
end of the jet. Possible intraknot faint X-ray emission is present. 

\noindent {\bf 1150+497:} The \vla\ contours in Figure~1 were
restored with beamsize 0.5\arcsec. The lowest contour is 0.25
mJy/beam. The radio images were reprocessed from calibrated data
published in Owen \& Puschell (1984). The knots nomenclature changed
from Paper I, following the detection of several more optical
counterparts in the \hst\ data. The innermost optical knot A at
0.9\arcsec\ is not resolved at X-rays and we can only give an upper
limit in Table 3a.  Detected X-ray knot B at 2.2\arcsec\ (Table 2)
correponds to the sum of two optical knots at 2.1\arcsec\ and
2.6\arcsec. The X-ray detection of the hotspot in the Southern lobe (H 
in Table 2) is resolved in the optical into two distinct components,
one at 8.3\arcsec\ and one at 8.5\arcsec. Note the ``corkscrew''
structure of this jet, reminiscent of 3C~273.

\noindent {\bf 1354+195:} The \vla\ contours in Figure~1 were
restored with beamsize 1\arcsec. The lowest contour is 0.6
mJy/beam. Our analysis is consistent with Paper I. However, due to the
criteria for detection adopted here, a few faint X-ray knots that were
considered detections in Paper I now do not meet the detection
criteria defined above, and are instead listed in Table 3a (knots C, D,
E, H). Solid detections at $2\sigma$ or more were obtained for the
remaining knots A, B, F, G, and the hotspot I in the Southern
lobe. Optical counterparts are present only for the two innermost
knots A and B. This is the longest (28\arcsec\ in projected size) and
narrowest jet of the sample. 

\noindent {\bf 1510$-$089:} The \vla\ contours in Figure~1 were
restored with beamsize 0.4\arcsec. The lowest contour is 0.45
mJy/beam. Three X-ray knots are detected (Table 2), with no optical
counterparts. This jet has an interesting morphology. The inner part
up to knot B follows the radio closely, with an almost one-to-one
correspondence. After knot B, the jet widens at both X-rays and radio,
and there is no clear correspondence between the two wavelengths. The
X-ray jet is shorter than at radio: X-ray emission ends at $\sim$
5\arcsec\ from the core, while the radio continues up to $\sim$
10\arcsec, bending slightly to the West.

\noindent {\bf 1641+399:} The \vla\ contours in Figure~1 were
restored with beamsize 0.45\arcsec. The lowest contour is 4.0
mJy/beam. There is only one X-ray knot detected at 2.7\arcsec\ N-W
from the core, which also has an optical counterpart. The X-ray jet is
shorter than the radio one.

\noindent {\bf 1642+690:} The \vla\ contours in Figure~1 were
restored with beamsize 0.45\arcsec. The lowest contour is 0.4
mJy/beam. X-ray emission is clearly detected in a knot at 2.7\arcsec\
from the core, with no optical counterpart. Optical emission is
instead present from the inner knot at 0.7\arcsec\ (Table 3). This is
another example of a jet shorter at X-rays than in the radio.

\noindent {\bf 1741+279:} The \vla\ contours in Figure~1 were
restored with beamsize 0.4\arcsec. The lowest contour is 0.15
mJy/beam. This is an unclear case. Inspection of the \chandra\ map
shows possible faint emission from the inner jet around 1.8\arcsec,
roughly at the center of the elongated radio feature. However, a
similar feature is present at the opposite azimuth. There is a very
weak detection of the N hotspot which we list in Table~3b.

\noindent {\bf 1928+738:} The \vla\ contours in Figure~1, taken at 1.4
GHz, were restored with beamsize 1.5\arcsec. The lowest contour is
0.75 mJy/beam. Only one X-ray knot is detected at 2.6\arcsec\ from the
nucleus, which also has an optical counterpart. This is also the only
well-defined radio feature, after which the radio jet widens into a
poorly confined lobe.

\noindent {\bf 2251+134:} The \vla\ contours in Figure~1 were
restored with beamsize 0.45\arcsec. The lowest contour is 0.95
mJy/beam. Our \hst\ image shows optical emission from three knots at
1.1\arcsec\ (A), 2.3\arcsec, and 2.8\arcsec.  We list an upper limit
to the innermost knot, A, in Table 3a.




\begin{figure}[h]
\noindent{\psfig{file=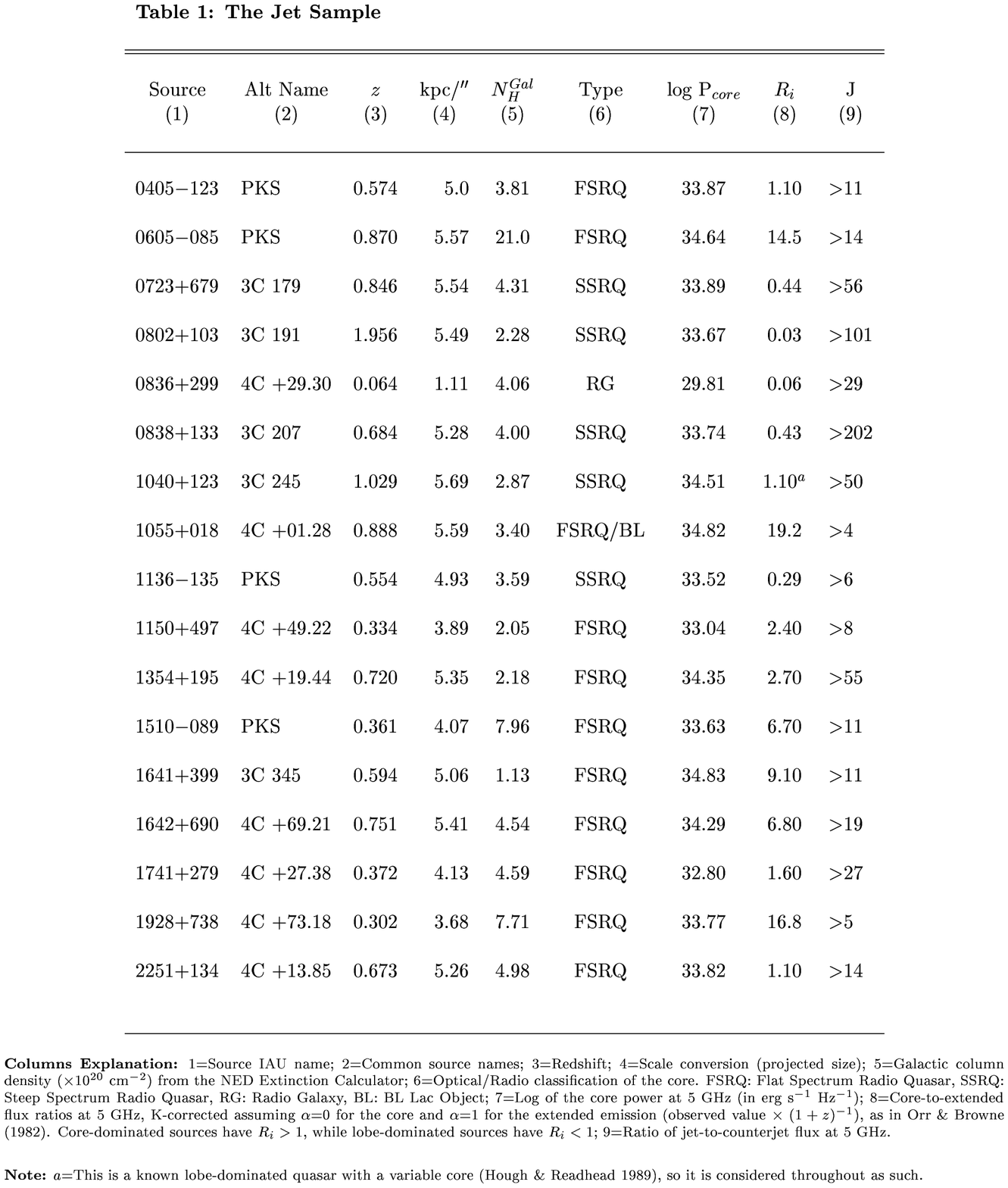,width=17cm,height=25cm,angle=0}}
\end{figure}

\begin{figure}[]
\noindent{\psfig{file=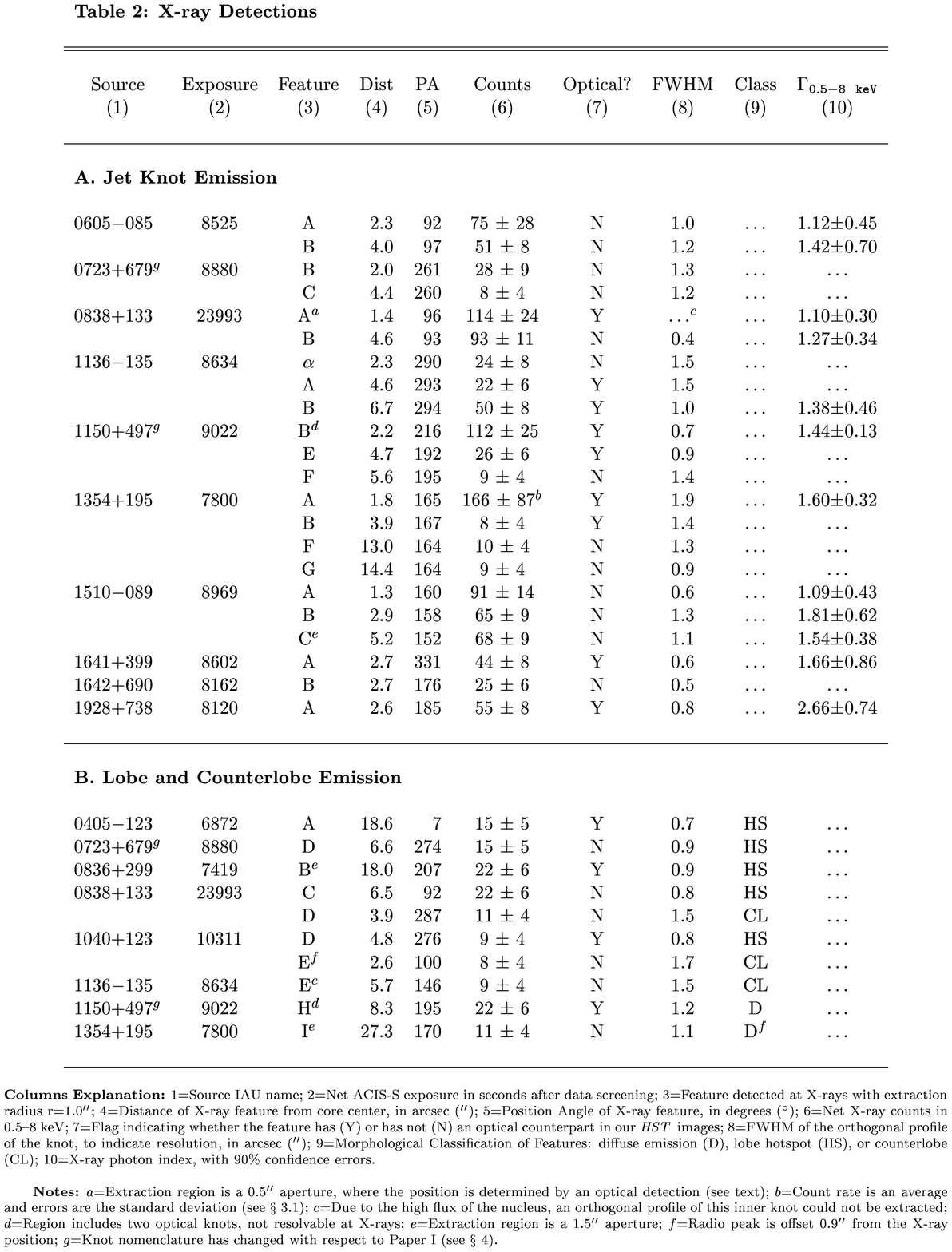,width=17cm,height=25cm,angle=0}}
\end{figure}

\begin{figure}[]
\noindent{\psfig{file=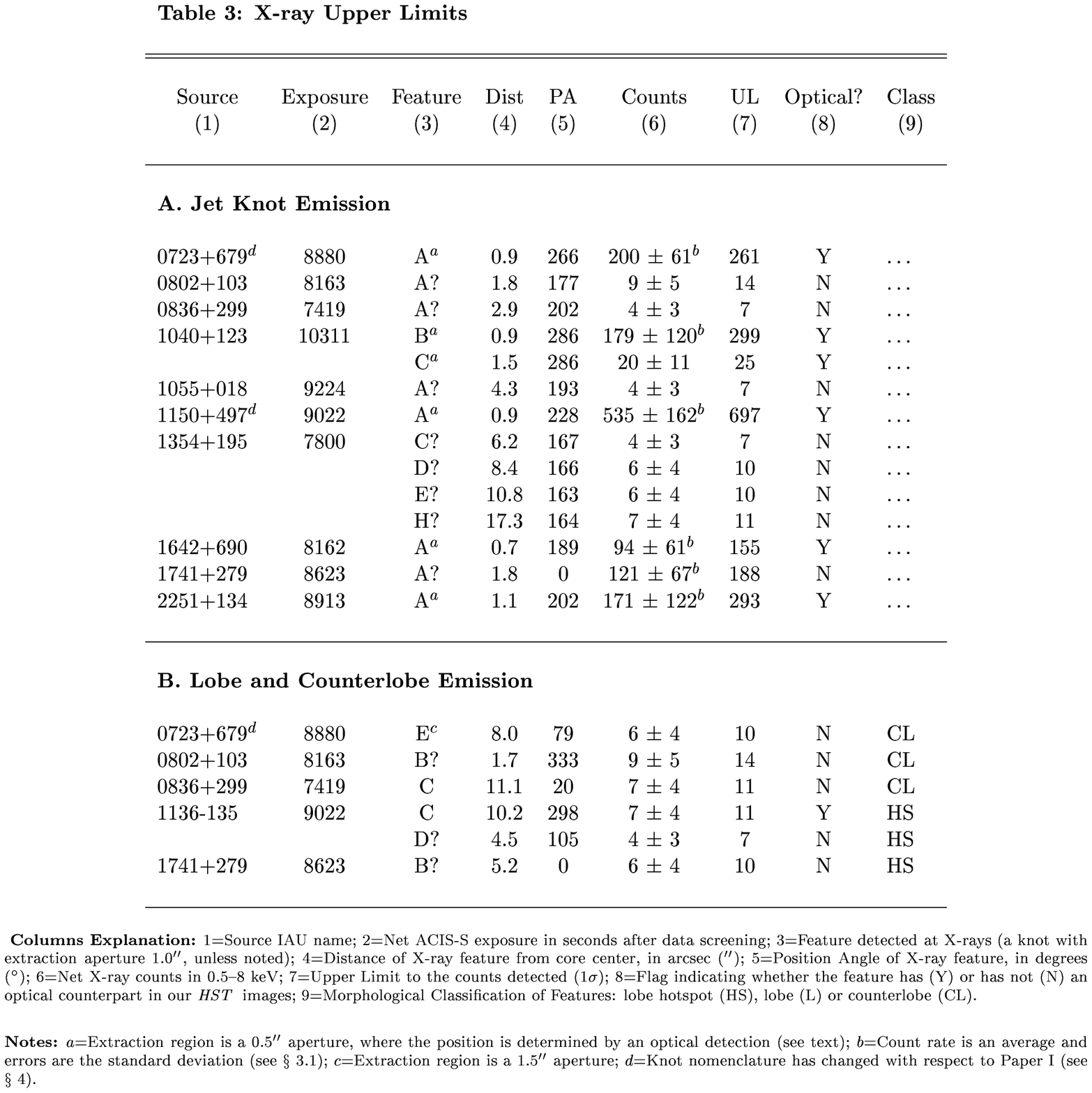,width=17cm,height=25cm,angle=0}}
\end{figure}

\begin{figure}[]
\noindent{\psfig{file=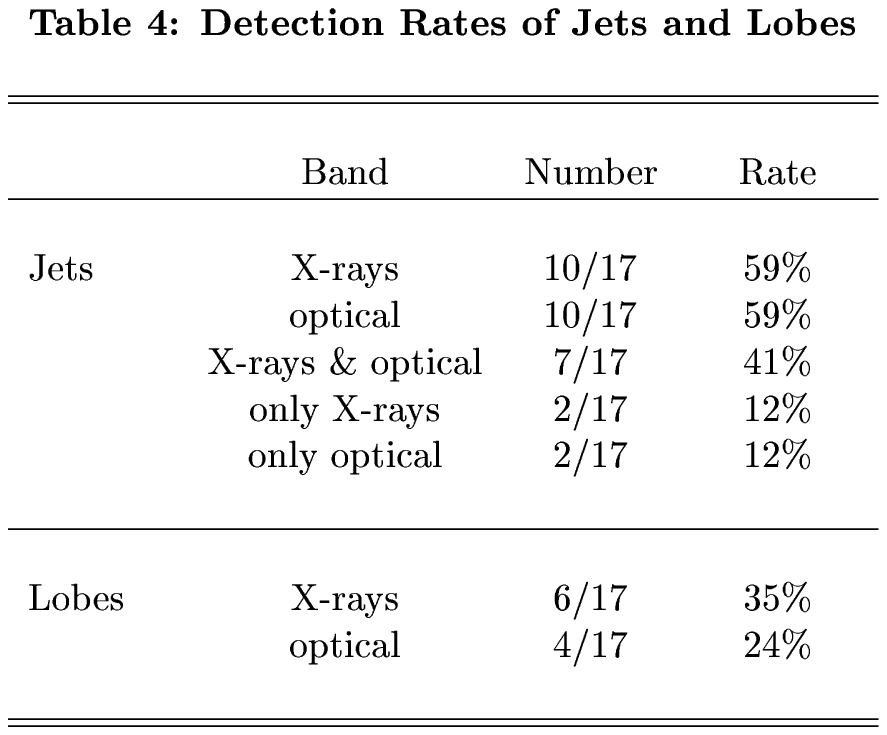,width=17cm,height=25cm,angle=0}}
\end{figure}

\begin{figure}[]
\noindent{\psfig{file=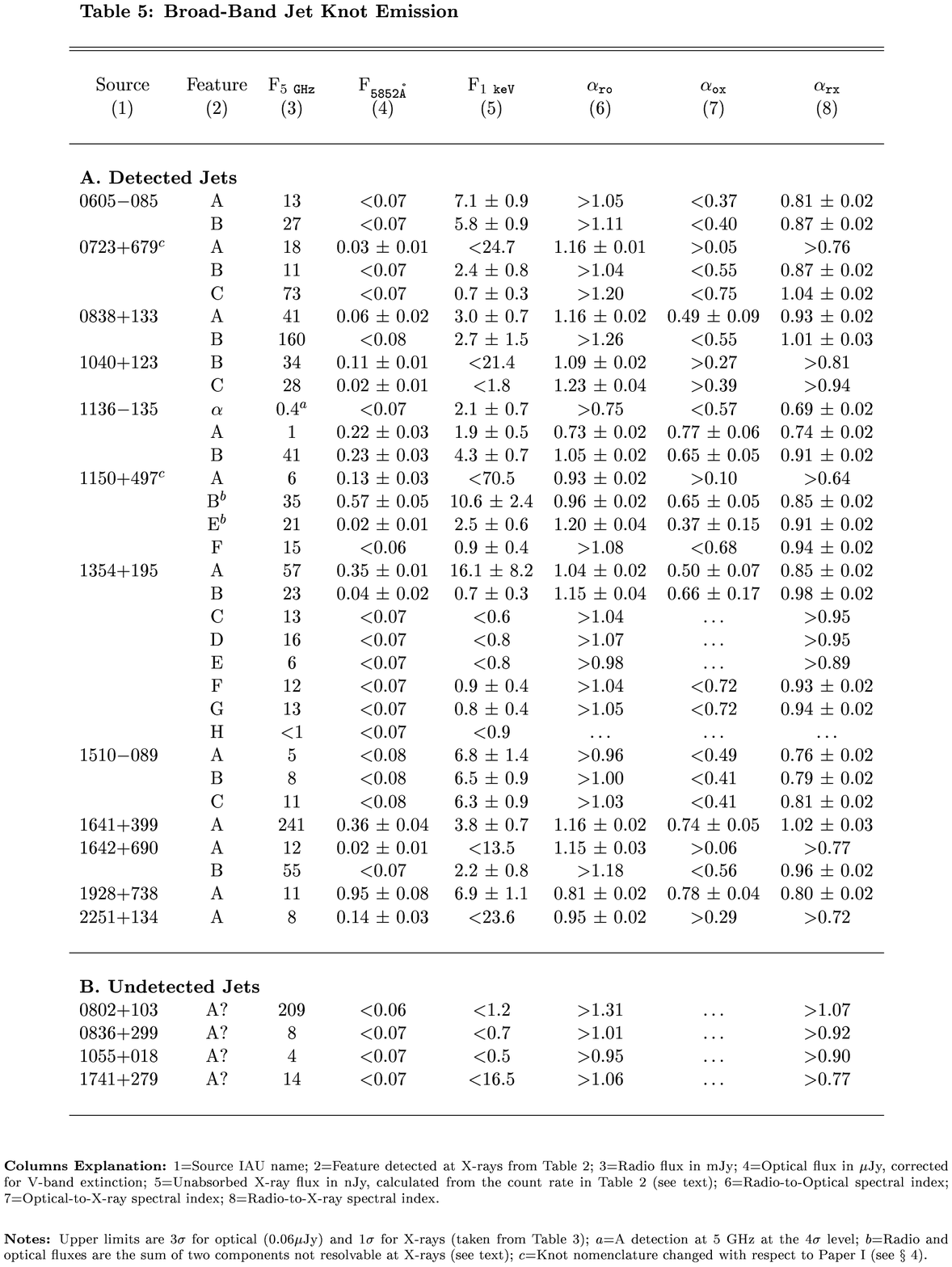,width=17cm,height=25cm,angle=0}}
\end{figure}

\begin{figure}[]
\noindent{\psfig{file=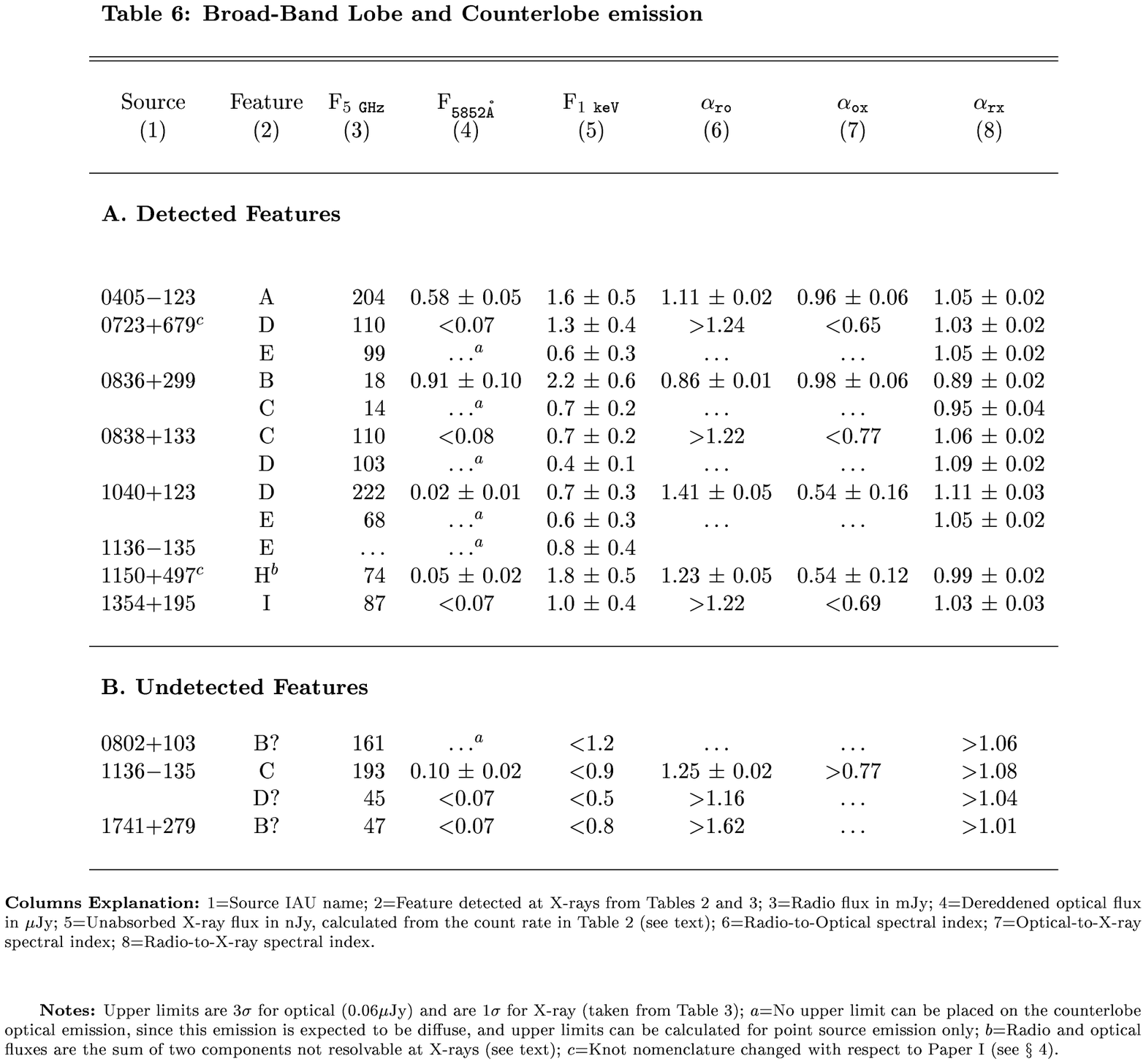,width=17cm,height=25cm,angle=0}}
\end{figure}

\begin{figure}[]
\noindent{\psfig{file=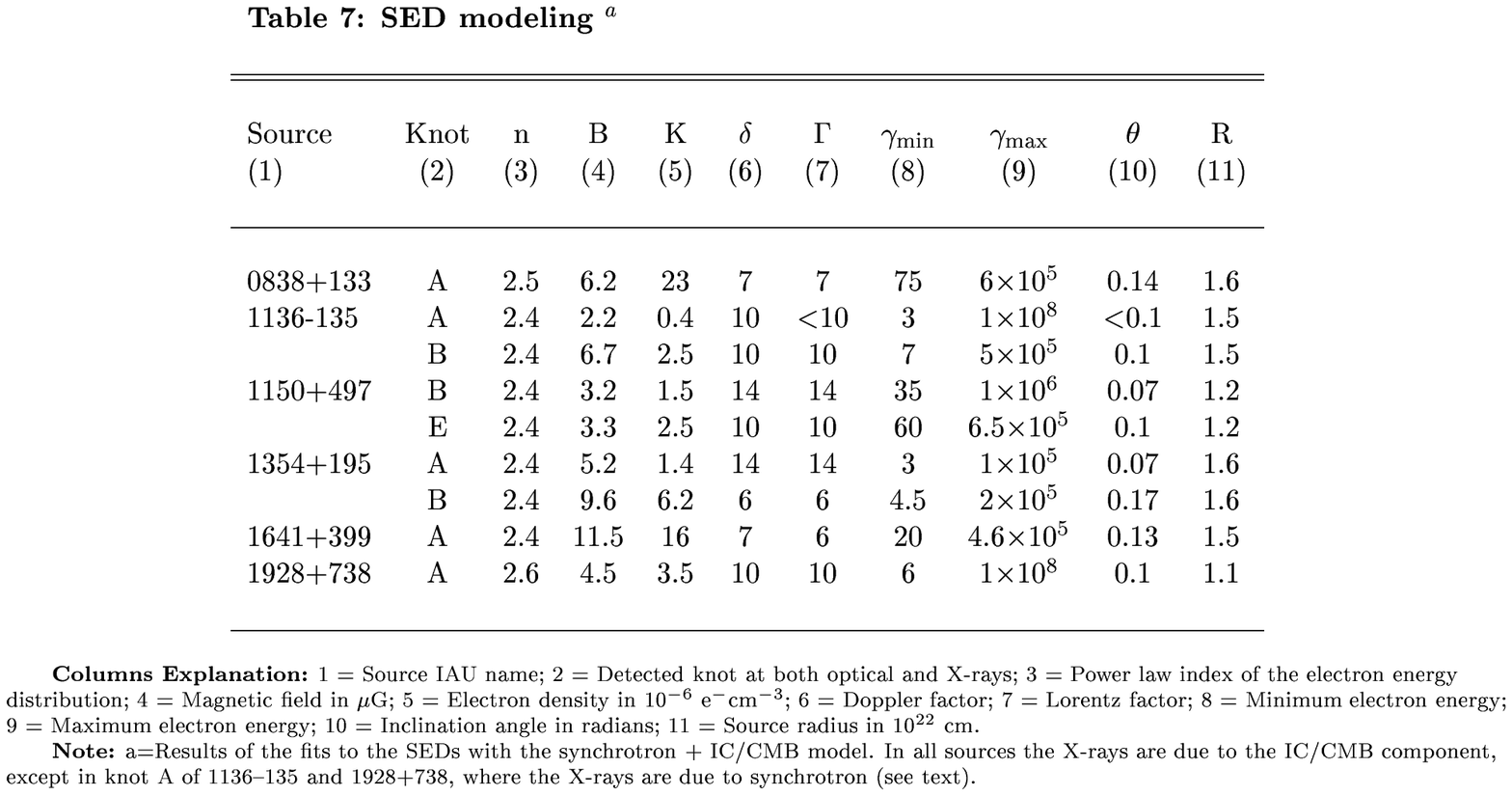,width=17cm,height=25cm,angle=0}}
\end{figure}


\begin{figure}
\noindent \caption{\chandra\ ACIS-S images of the radio jets of our
sample in the 0.5--8 keV energy range.  Overlaid are the radio
contours from archival \vla\ data. Both the colors and the contours
are plotted logarithmically, in steps of factor 2. The \chandra\ image
is smoothed with a Gaussian of width $\sigma$=0.3\arcsec, yielding a
resolution of 0.86\arcsec\ FWHM.  The radio image was restored with a
circular beam (see Appendix). North is up and East to the left.}
\end{figure}





\begin{figure}[]
\caption{\chandra\ ACIS-S images in the 0.5--8 keV energy
range of selected jets of our sample, showing the innermost jet
regions close to the core. North is up and East to the left.}
\end{figure}

\begin{figure}[]
\noindent{\psfig{file=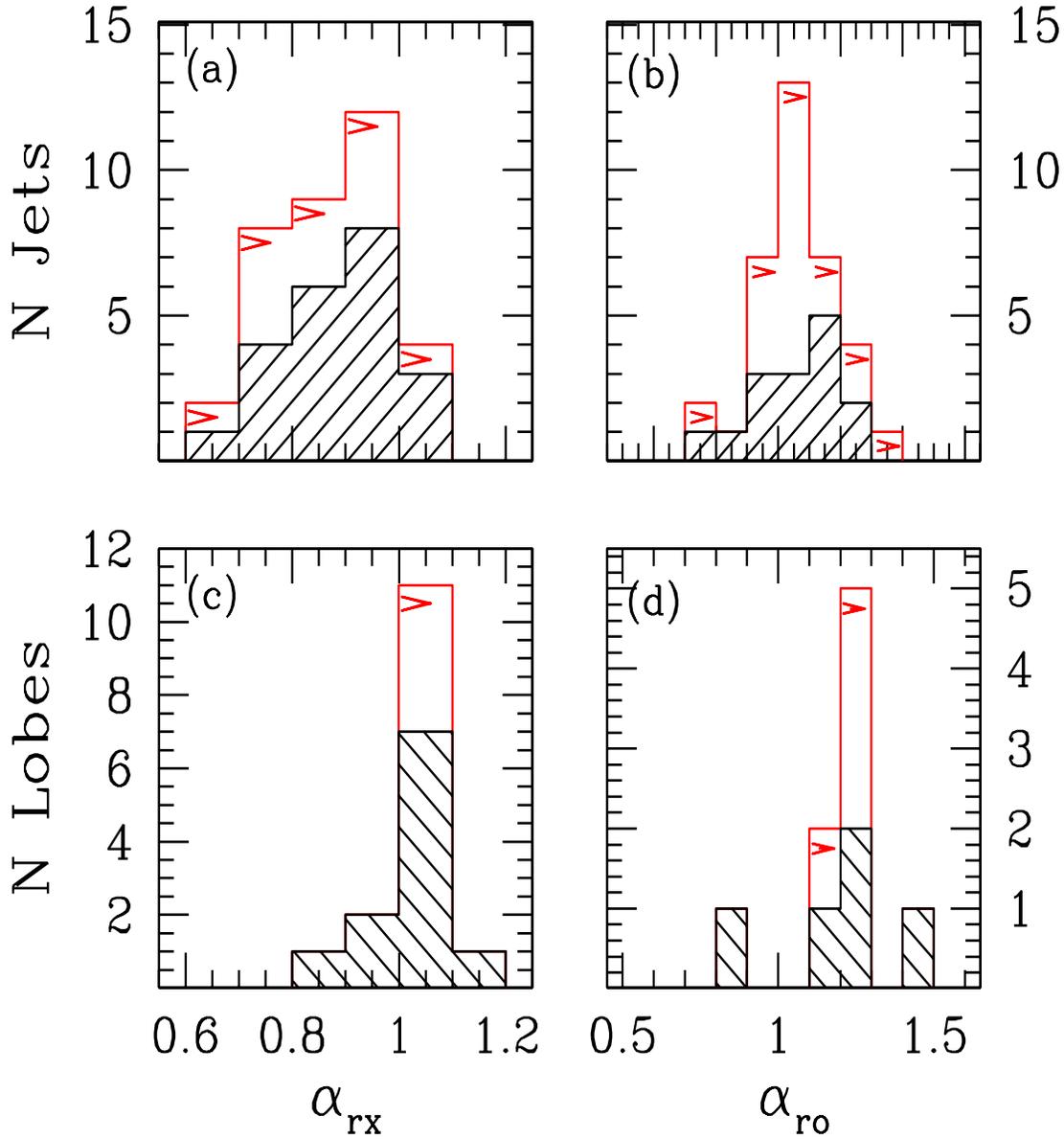,width=17cm,height=20cm,angle=0}}
\caption{Distributions of the radio-to-X-ray spectral index,
$\alpha_{rx}$ (left panels) and for the radio-to-optical index, 
$\alpha_{ro}$ (right panels). {\it (a) and (b):} Distributions of
$\alpha_{rx}$ and $\alpha_{ro}$ for the jet knots; {\it (c) and (d):} Distributions of
$\alpha_{rx}$ and $\alpha_{ro}$ for the lobes. The dashed histograms
represent detections, while the arrows represent lower limits.} 
\end{figure}

\begin{figure}[]
\noindent{\psfig{file=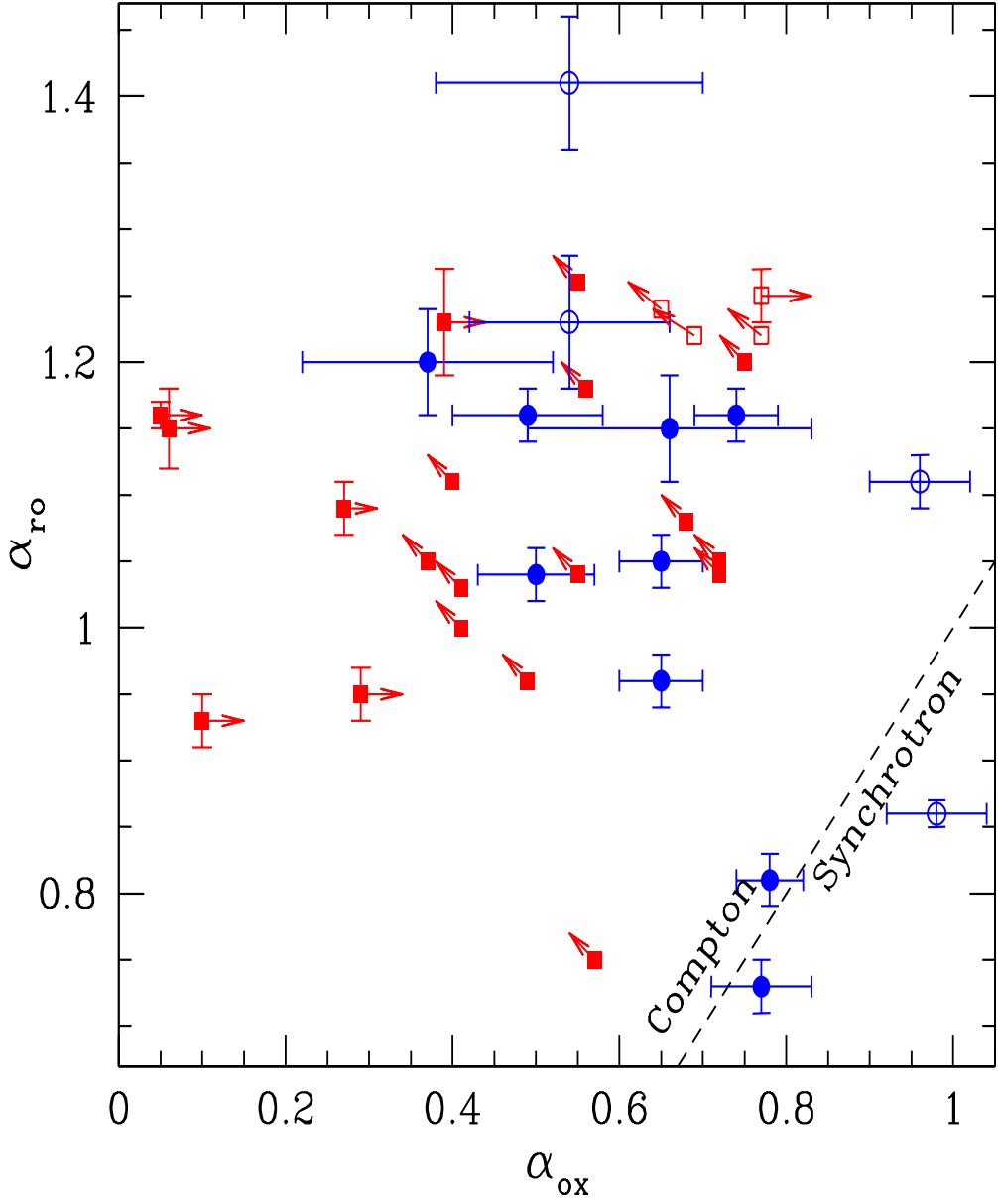,width=14cm,height=17cm,angle=0}}
\caption{Plot of the radio-to-optical spectral index, $\alpha_{ro}$,
versus the optical-to-X-ray index, $\alpha_{ox}$. The filled circles 
are for the jet knots (Table 5), while the open circles are for the
lobes (Table 6). The squares are for non-detections. 
The dashed line marks the locus where $\alpha_{ox}=\alpha_{ro}$. Knots
lying above the dashed line have a concave SED, where the X-rays
belong to a different spectral component than the longer wavelengths;
IC probably dominates in these sources for the X-ray production. Knots
below the dashed line have a convex SED, and the X-rays are
interpreted as due to synchrotron emission (see text).}
\end{figure}

\begin{figure}[]
\noindent{\psfig{file=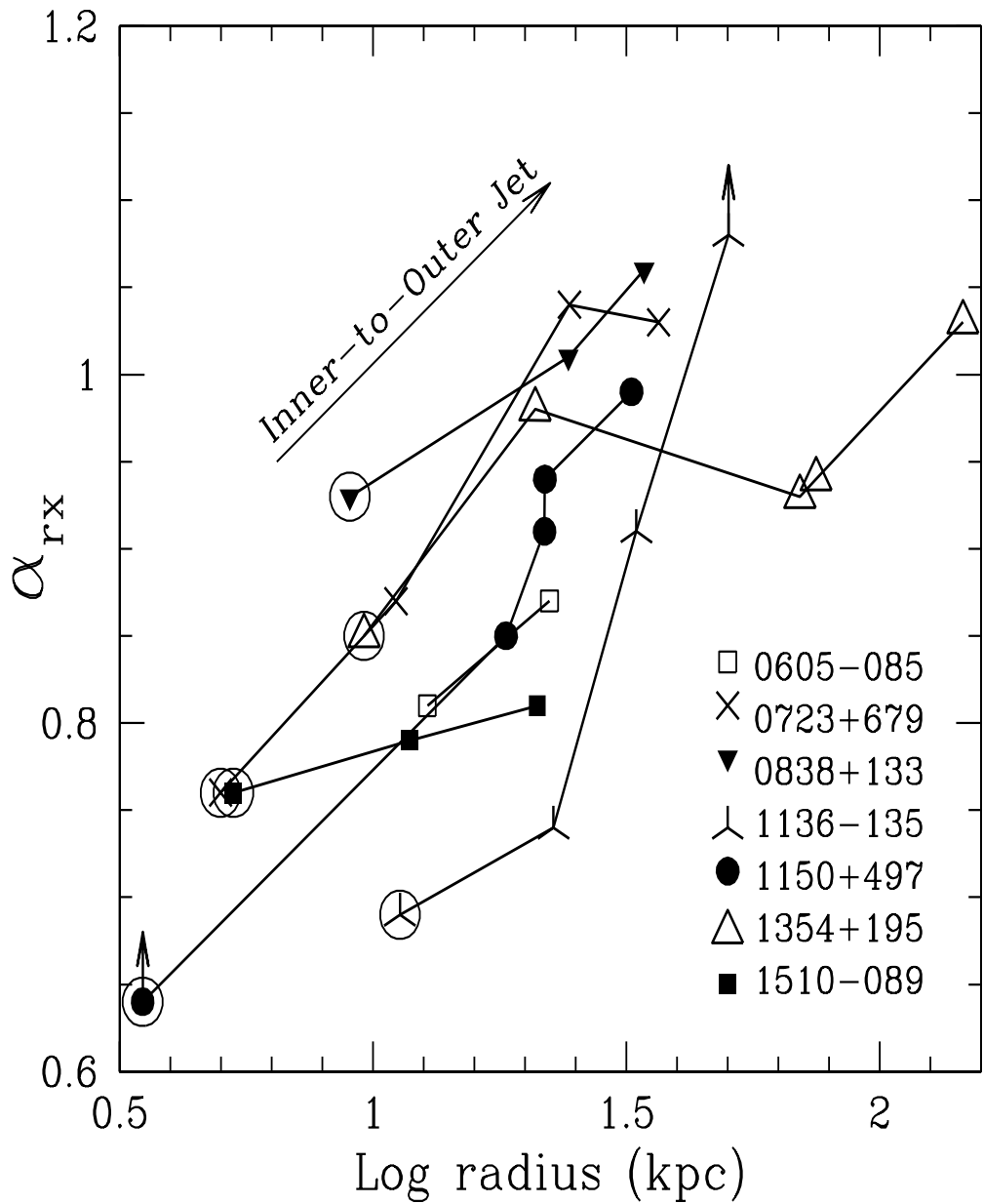,width=14cm,height=17cm,angle=0}}
\caption{Plot of the radio-to-X-ray spectral index for the knots of
individual jets (in different symbols) versus the projected distance
of the knot from the core. The circles mark the innermost knots where
contamination from the PSF wings could be present. The uncertainties
on $\alpha_{rx}$ are listed in Table 5 and 6. Interestingly, the
X-ray-to-radio flux ratio decreases with increasing distance along the
jet.}
\end{figure}

\begin{figure}[]
\noindent{\psfig{file=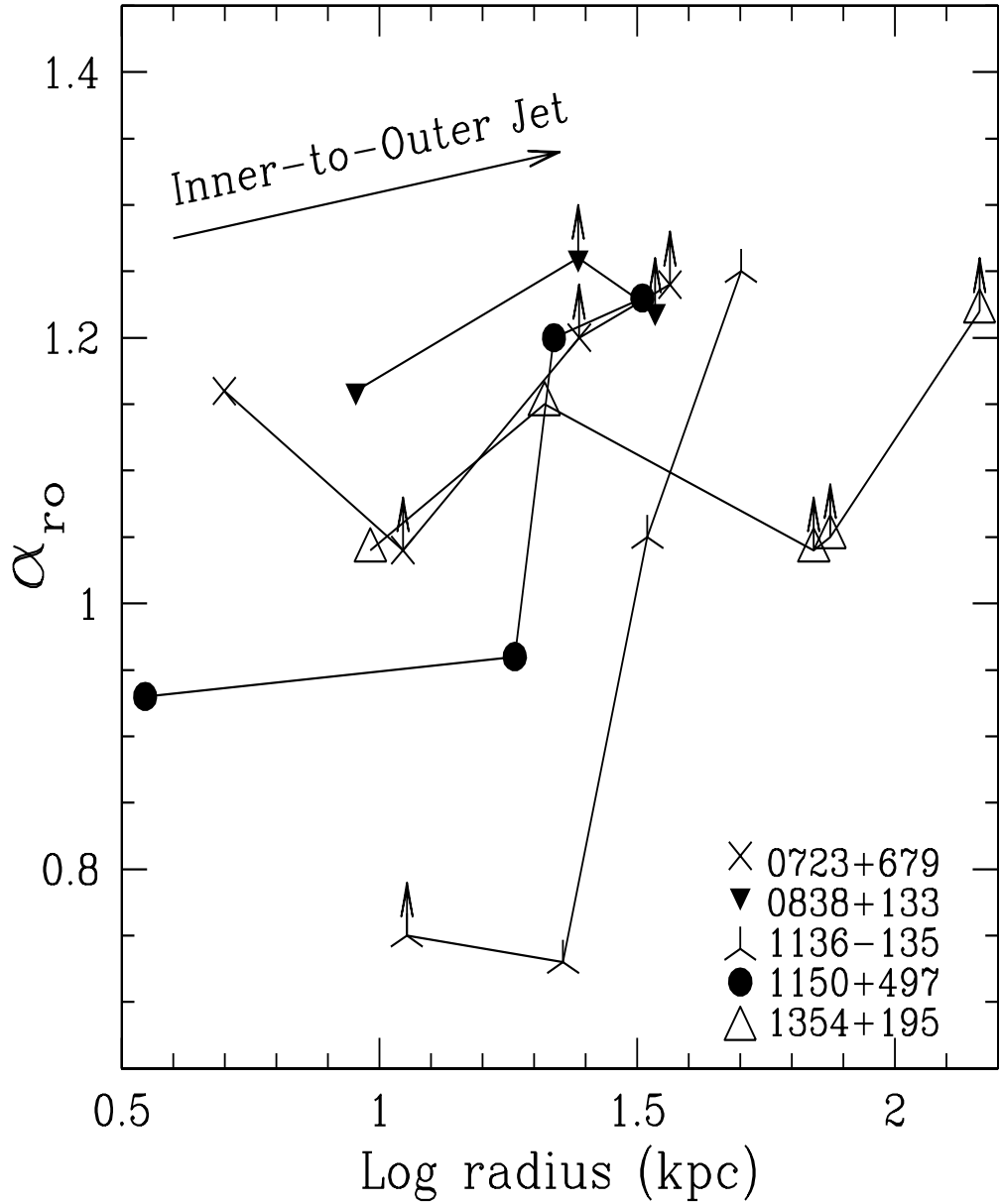,width=14cm,height=17cm,angle=0}}
\caption{Same as for Figure 5, but for the radio-to-optical index. 
Again, the optical-to-radio flux ratio decreases with increasing
distance along the jet.}
\end{figure}

\begin{figure}[]
\noindent{\psfig{file=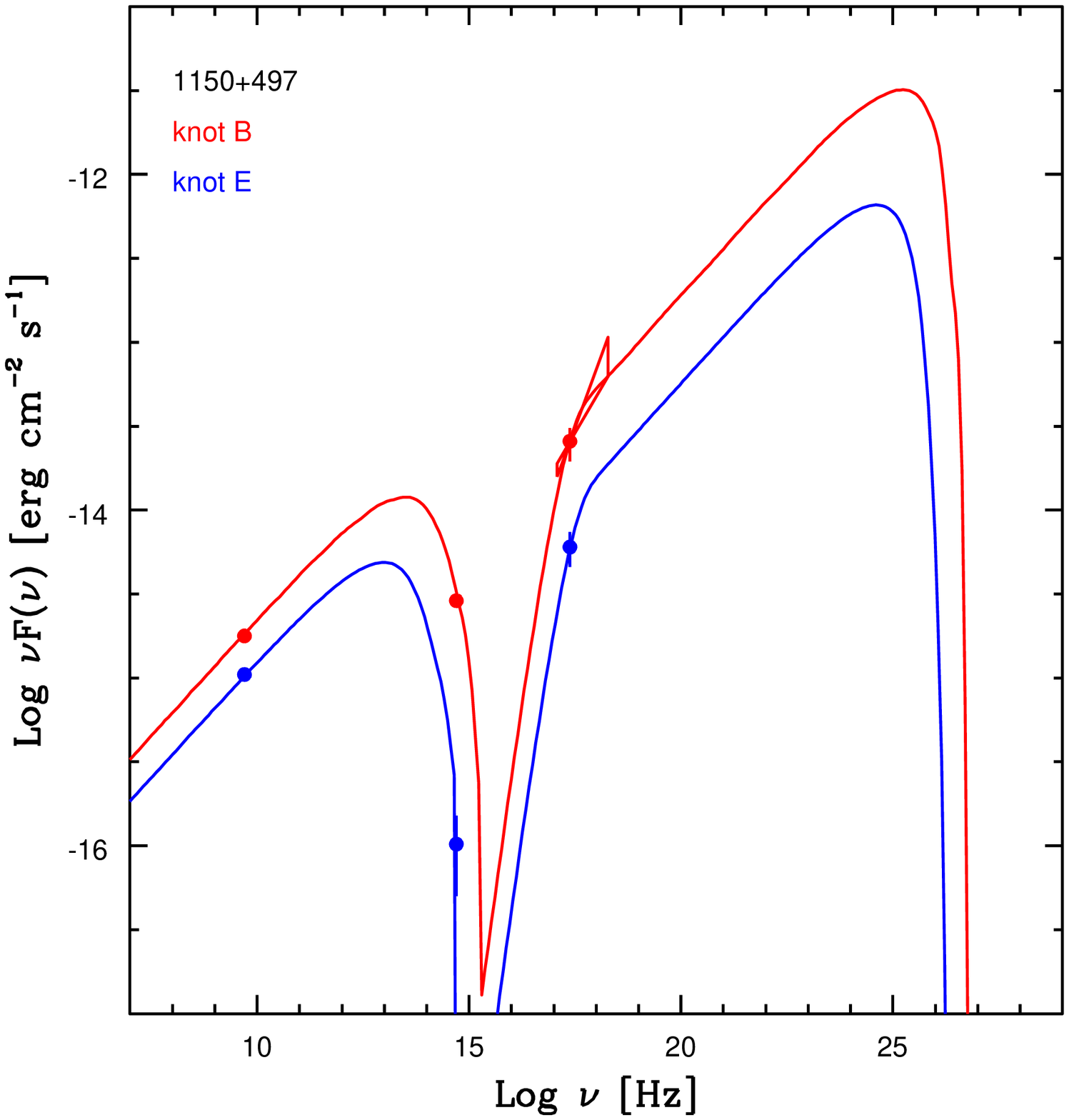,width=8cm,height=8cm,angle=0}}
{\psfig{file=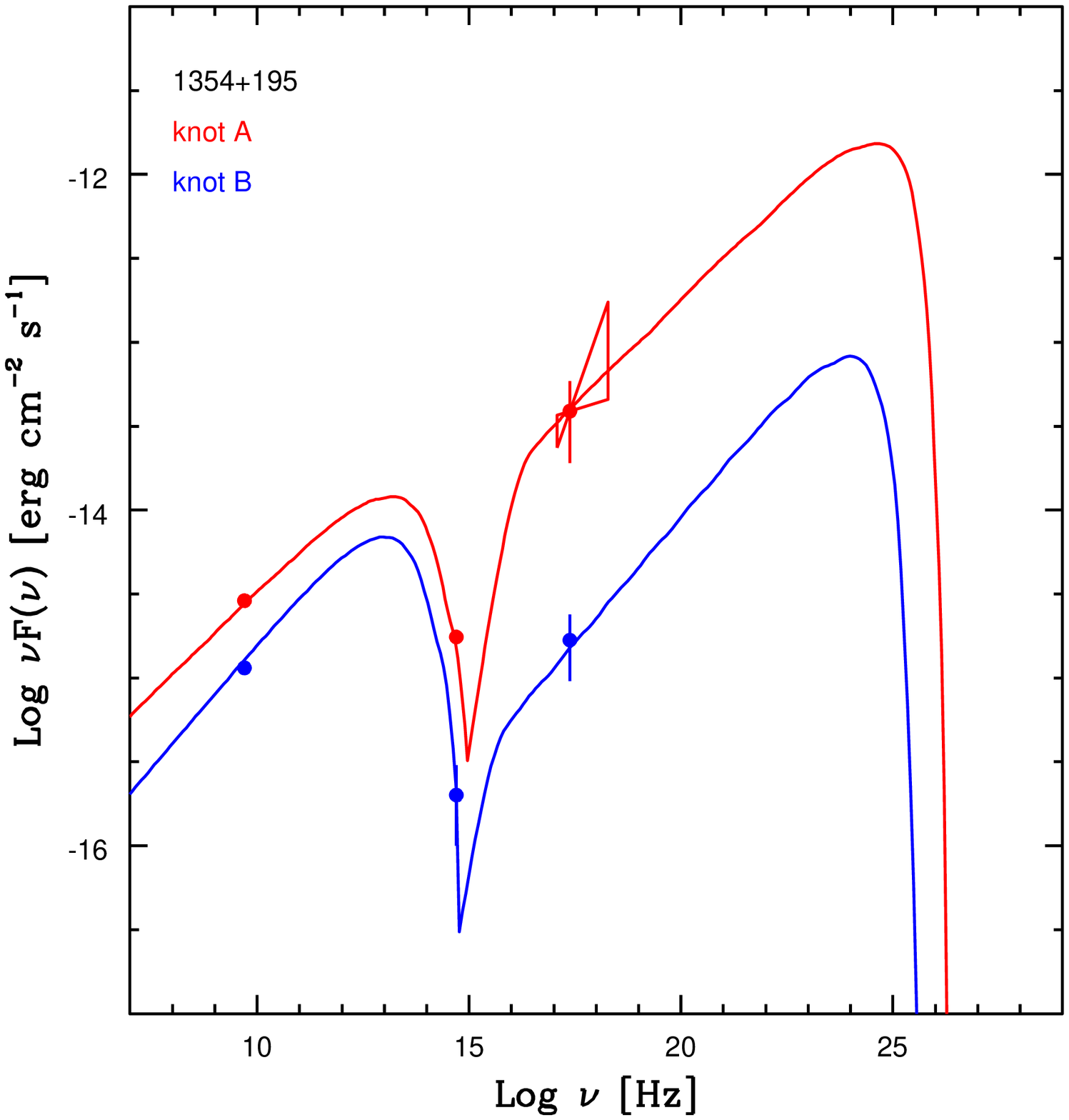,width=8cm,height=8cm,angle=0}}
\caption{Selected radio-to-X-ray Spectral Energy Distributions
(SEDs) and best-fit models assuming a synchrotron + IC/CMB model, with
the parameters reported in Table 7.} 
\end{figure}

\begin{figure}[]
\noindent{\psfig{file=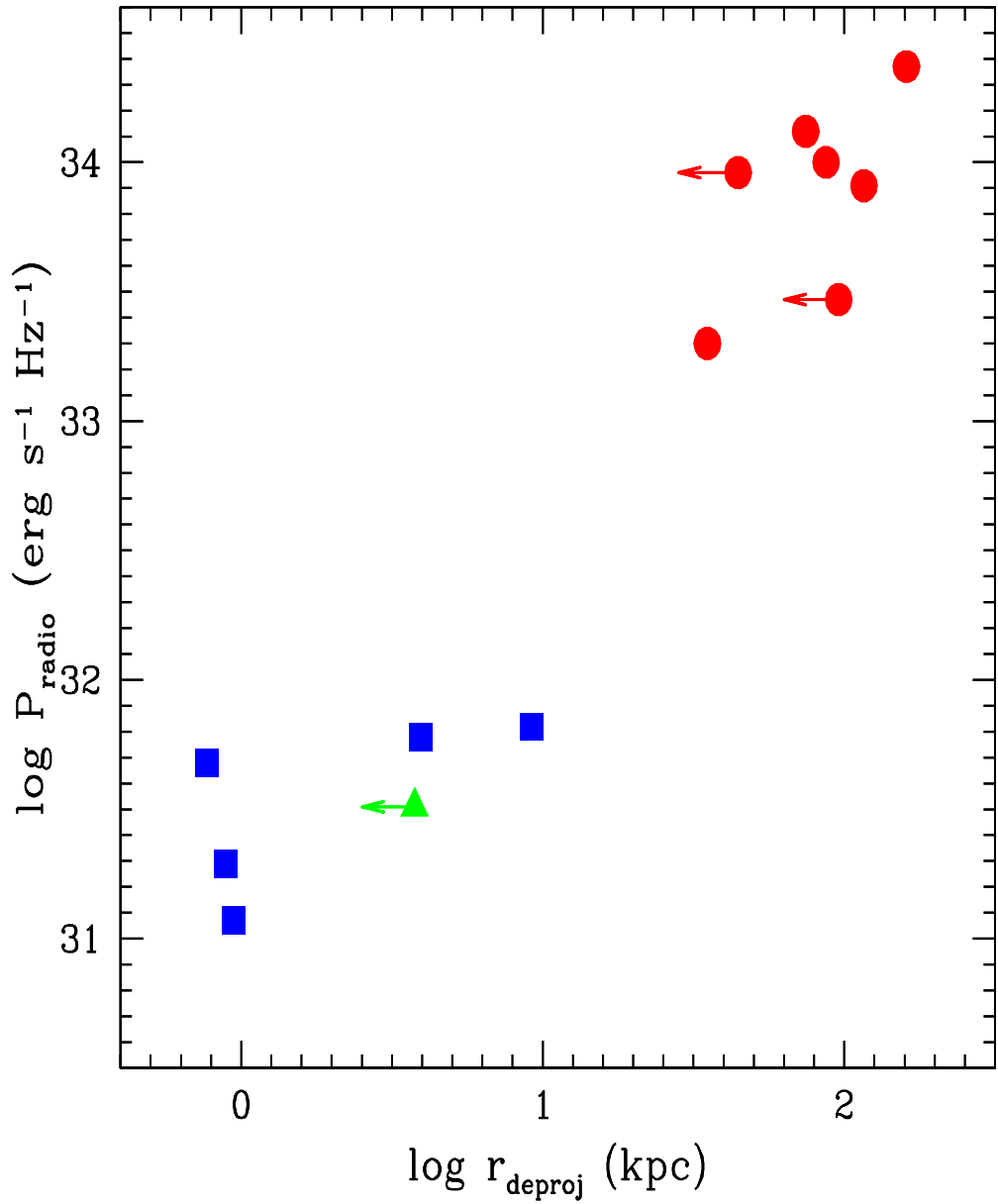,width=14cm,height=17cm,angle=0}}
\caption{Plot of the total radio power versus the {\it deprojected}
distance of the first detected optical/X-ray knot. The circles are the
FRIIs of our sample, while the squares are FRI sources from the
literature for which sufficient information is available (see
text). The triangle is the only FRI of our sample, 0836+299. In more
powerful sources, the first detected knot is at larger distances from
the core than in lower-power sources.} 
\end{figure}

\end{document}